\documentclass[preprint]{aastex}

\shorttitle{Cosmology with Strong Lensing Systems}
\shortauthors{Cao et al.}

\begin{document}


\title{Cosmology with Strong Lensing Systems}


\author{Shuo Cao}
\affil{Department of Astronomy, Beijing Normal University,
    Beijing 100875, China}

\author{Marek Biesiada}
\affil{Department of Astronomy, Beijing Normal University,
    Beijing 100875, China; \\
    Department of Astrophysics and Cosmology, Institute of Physics, University of Silesia, Uniwersytecka 4, 40-007, Katowice, Poland}

\author{Rapha\"{e}l Gavazzi}
\affil{Institute d'Astrophysique de Paris, UMR7095 CNRS - Universite Pierre et Marie Curie, 98bis bd Arago, 75014 Paris, France}

\author{Aleksandra Pi{\'o}rkowska}
\affil{Department of Astrophysics and Cosmology, Institute of Physics, University of Silesia, Uniwersytecka 4, 40-007, Katowice, Poland}

\and

\author{Zong-Hong Zhu}
\affil{Department of Astronomy, Beijing Normal University,
    Beijing 100875, China}
\email{zhuzh@bnu.edu.cn}



\begin{abstract}
In this paper, we assemble a catalog of 118 strong gravitational
lensing systems from SLACS, BELLS, LSD and SL2S surveys and use them
to constrain the cosmic equation of state. In particular we consider
two cases of dark energy phenomenology: $XCDM$ model where dark
energy is modeled by a fluid with constant $w$ equation of state
parameter and in Chevalier - Polarski - Linder (CPL) parametrization
where $w$ is allowed to evolve with redshift: $w(z) = w_0 + w_1
\frac{z}{1+z}$. We assume spherically symmetric mass distribution in
lensing galaxies, but relax the rigid assumption of SIS model in
favor to more general power-law index $\gamma$, also allowing it to
evolve with redshifts $\gamma(z)$. Our results for the $XCDM$
cosmology show the agreement with values (concerning both $w$ and
$\gamma$ parameters) obtained by other authors. We go further and
constrain the CPL parameters jointly with $\gamma(z)$.
The resulting confidence
regions for the parameters are much better than those obtained with
a similar method in the past. They are also showing a trend of being
complementary to the supernova Ia data. Our analysis demonstrates
that strong gravitational lensing systems can be used to probe
cosmological parameters like the cosmic equation of state for dark
energy. Moreover, they have a potential to judge whether the cosmic
equation of state evolved with time or not.

\end{abstract}





\section{Introduction}
Significance of strong gravitational lensing in cosmology has been
recognized quite early. Seminal work of \citet{Refsdal64}
demonstrating possibilities of independent determination of the
Hubble constant from time delays between images gave momentum to the
development of strong gravitational lensing theory. However, it has
only recently been possible to determine the Hubble constant from
lensing time delays with the precision competitive with other
techniques \citep{Suyu10}. Currently, however, goals and performance
of cosmology go far beyond ``the quest for two numbers'' ($H_0$ and
$q_0$ -- the deceleration parameter). With discovery of the
accelerating expansion of the Universe \citep{Riess98,Perlmutter99},
understanding this phenomenon has become one of the most important
issues of modern cosmology. Because there is still no fully
convincing hints from the side of the theory, we are left with a
pragmatic approach to model this phenomenon as the so called dark
energy ---  hypothetical homogeneous fluid phenomenologically
described by the barotropic equation of state $p = w \rho$. This
approach provides a particularly suitable link between theory and
observations. Namely, it encompasses the case of cosmological
constant $\Lambda$ ($w=-1$) and that of the scalar field. If the
scalar field settled in an attractor \citep{Ratra88}, $w$ can now be
constant (with the fundamental demand that $w< -1/3$ in order to get
the acceleration) --- this is the so-called XCDM cosmology. It
contains $\Lambda$CDM as its special case and indeed all constraints
on $w$ obtained up to now within XCDM cosmology gave $w$ very close
to $-1$. Therefore now there is no debate whether the Universe is
accelerating, but rather whether $w$ coefficient evolved in time in
quite recent epochs (say at $z = 1 - 6$). It is a reasonable
question because -- as lucidly remarked by Linder -- if the scalar
field stands behind accelerating expansion, it should have evolved
since the only other case of the scalar field we know (the inflaton)
clearly evolved because the inflation ended. A very convenient
parametrization of $w(z)$ has been proposed by
\citet{Chevalier01,Linder03}: $w(z) = w_0 + w_1 \frac{z}{1+z}$,
which is essentially the first-order (linear) Taylor expansion in
the scale factor -- the true gravitational degree of freedom in FRW
cosmology. We will denote this case as CPL thereafter. The problem
is that if we use all conventional cosmological probes like SNIa,
BAO or CMBR acoustic peaks, we are facing the degeneracy
(colinearity) between $w_0$ and $w_1$ parameters (stemming from the
fact that $w(z)$ overall should be negative). Standard rulers (like
BAO and CMBR) and standard candles (SNIa) when used to constrain
$\Omega_m$ and $w$ parameters acted complementarily in the sense
that their respective confidence regions in $(\Omega_m,w)$ plane
were almost orthogonal. No such complementarity has been proposed so
far for $w_0$, $w_1$ parameters, although it has been noticed by
\citet{Linder04} that strong lensing systems are promising in this
respect. It is one of the most important reasons motivating our
work. Earlier attempts to use strong lensing systems for
constraining parameters of the cosmological model were based on two
approaches. First was the statistical one based on comparison
between empirical distribution of image separations in observed
samples of lenses and the theoretical one --- using CLASS
\citep{Chae02}, or SQLS samples \citep{Oguri12} respectively. Second
approach made use of galaxy clusters in the role of lenses
\citep{Paczynski81,Sereno02,Meneghetti05,Gilmore09,Jullo10} where a
single lens (cluster) typically generates ca. 100 images of ca.
30-40 different sources (distant galaxies).

We use another method which can be traced back to the papers of e.g.
\citet{Futamase01,Biesiada06,Grillo08}. However, similarly as with
the Hubble constant, it is only quite recently when reasonable
catalogs of strong lenses: containing more than 100 lenses, with
spectroscopic as well as astrometric data, obtained with well
defined selection criteria are becoming available. It is also only
recently when our knowledge about structure and evolution of early
type galaxies allows us to undertake the assessment of such important
factor as mass density profile. Recent works
\citep{Biesiada10,Cao12} provided first successful applications of
the proposed method in the context of dark energy, although small
samples have not allowed for stringent constraints. This encourages
us to improve and develop it further. On a new sample of 118 lenses
compiled from SLACS, BELLS, LSD and SL2S surveys we constrain the
cosmic equation of state in XCDM cosmology and in CPL
parametrization. Moreover we relax the rigid assumption of SIS model
to general power-law density profile allowing the power-law index to
evolve with redshifts.

This paper is organized as follows. In Section~\ref{sec:method}, we
briefly describe the methodology. Then, in Section~\ref{sec:data} we
introduce the strong lensing data used in our analysis. The results
are presented in Section~\ref{sec:results} and concluded in
Section~\ref{sec:conclusions}.

\section{Methodology} \label{sec:method}

As one of the successful predictions of General Relativity in the
past decades, strong gravitational lensing has become a very
important astrophysical tool allowing us to use individual lensing
galaxies to measure cosmological parameters
\citep{Treu06,Grillo08,Biesiada10,Cao12}. For a specific strong
lensing system with the intervening galaxy acting as a lens, the
multiple image separation of the source depends only on angular
diameter distances to the lens and to the source, as long as one has
a reliable model for the mass distribution within the lens.

Moreover, compared with late-type and unknown-type counterparts,
early-type galaxies (ETGs, ellipticals) are more likely to serve as
intervening lenses for the background sources (quasars or galaxies).
This is because such galaxies contain most of the cosmic stellar
mass of the Universe. This property affects statistics of
gravitational lensing phenomenon which leads to a sample dominated
by early-type galaxies (see \citet{Kochanek00} and references
therein). Here, we encounter a critical issue that still needs to be
addressed in a systematic way: such properties of early-type
galaxies like their formation and evolution are still not fully
understood. Fortunately, a sample of well-defined strong lensing
systems is now available and can be used to study the mass density
distribution in early-type galaxies.

In principle, the lens model often fitted to the observed images is
based on a singular isothermal ellipsoid (SIE) model, in which the
projected mass distribution is elliptical \citep{Ratnatunga99}.  In
this paper, we will take a simpler approach and assume spherical
symmetry. For a moment we will refer to the singular isothermal
sphere (SIS) model instead which will then be generalized.

The main idea of our method is that formula for the
Einstein radius in a SIS lens
\begin{equation} \label{E radius}
 \theta_E = 4 \pi
\frac{\sigma_{SIS}^2}{c^2} \frac{D_{ls}}{D_s}
\end{equation}
depends on the cosmological model through the ratio of
(angular-diameter) distances between lens and source and between
observer and lens. The angular diameter distance in flat
Friedmann-Robertson-Walker cosmology reads
\begin{equation} \label{angular}
D(z;{\mathbf p}) = \frac{1}{1+z} \frac{c}{H_0}  \int_0^z
\frac{dz'}{h(z';{\mathbf p})}
\end{equation} where $H_0$ is the present value
of the Hubble function and $h(z;{\mathbf p})$ is a dimensionless
expansion rate dependent on redshift $z$ and cosmological model
parameters are: ${\mathbf p}=\{\Omega_m, w \}$ for $XCDM$ cosmology or
${\mathbf p}=\{\Omega_m, w_0,  w_1 \}$ for the CPL parametrization
of evolving equation of state. More specifically, $h^2(z;{\mathbf
p}) = \Omega_m (1+z)^3 + (1 - \Omega_m)(1+z)^{3(1+w)}$ for $XCDM$
cosmology and $h^2(z;{\mathbf p})= \Omega_m
(1+z)^3+(1-\Omega_m)(1+z)^{3(1+w_0+w_1)} \exp{(- \frac{3 w_1
z}{1+z}) }$ for CPL parametrization.

Provided one has a reliable knowledge about the lensing system, i.e.
the Einstein radius $\theta_E$ (from image astrometry) and stellar
velocity dispersion $\sigma_{SIS}$ (form central velocity dispersion
$\sigma_0$ obtained from spectroscopy), one can use it to test the
background cosmology. This method is independent of the Hubble
constant value (which gets canceled in the distance ratio) and is
not affected by dust absorption or source evolutionary effects. It
depends, however, on the reliability of lens modeling (e.g. SIS
assumption) and measurements of $\sigma_0$. Hopefully, starting with
the Lens Structure and Dynamics (LSD) survey and the more recent
SLACS survey, spectroscopic data for central parts of lens galaxies
became available allowing to assess their central velocity
dispersions. There is a growing evidence for homologous structure of
late type galaxies \citep{Koopmans06,Koopmans09,Treu06} supporting
reliability of SIS assumption.

In our method, cosmological model enters not through a distance measure directly, but rather through a distance ratio
\begin{equation} \label{observable}
{\cal D}^{th}(z_l,z_s; {\mathbf p}) = \frac{D_{ls}({\mathbf
p})}{D_{s}({\mathbf p})} = \frac{ \int_{z_l}^{z_s} \frac{dz'}{
h(z';{\mathbf p})}} {\int_{0}^{z_s} \frac{dz'}{ h(z';{\mathbf p})}}
\end{equation}
and respective observable counterpart reads
$$
{\cal D}^{obs} = \frac{c^2 \theta_E}{4 \pi \sigma_0^2}
$$

From one side this circumstance is a complication, but it also
offers an advantage. Namely, as we already mentioned, $w_0$ and
$w_1$ parameters are intrinsically anti correlated, which makes it
difficult to constrain them. In the ${\cal D} $ ratio, however a
competition between two angular diameter distances may lead to
positive correlations between $w_0$ and $w_1$ as discussed recently
in \citep{Piorkowska13}. Constraints on cosmological models using
this method have been obtained e.g. in
\citet{Biesiada10,Biesiada11,Cao12}.

One objection one might rise towards the proposed method is the
rigid assumption of the SIS model for the lens. Although there are
some arguments that inside Einstein radii total mass density follows
isothermal profile \citep{Koopmans06}, one can expect the deviation
from the isothermal profile and its evolution with redshift. The
last point theoretically stems from the structure formation theory
(ellipticals as a result of mergers) and to some extent has been
supported observationally \citep{Ruff11,Brownstein12,Sonnenfeld13b}.

Therefore we have to generalize the SIS model to spherically
symmetric power-law mass distribution $\rho \sim r^{- \gamma}$.
First, we recall that location of observed images, hence the
knowledge of $\theta_E$ provides us with the mass $M_{lens}$ inside
the Einstein radius: $M_{lens} = \pi R_E^2 \Sigma_{cr}$ where: $R_E
= \theta_E D_l$ is the physical Einstein radius (in [kpc]) in the
lens plane and $\Sigma_{cr} = \frac{c^2}{4 \pi G} \frac{D_s}{D_l
D_{ls}}$ is the critical projected mass density for lensing
\citep{Schneider92}. Finally we have:
\begin{equation} \label{lensing mass}
M_{lens} = \frac{c^2}{4 G} \frac{D_s D_l}{D_{ls}}\theta_E^2
\end{equation}
If one has spectroscopic data providing the velocity dispersion
$\sigma_{ap}$ inside the aperture (more precisely, luminosity
averaged line-of-sight velocity dispersion), then after solving
spherical Jeans equation (assuming that stellar and mass
distribution follow the same power-law and anisotropy vanishes) one
can assess the dynamical mass inside the aperture projected to lens
plane \citep{Koopmans05} and scale it to the Einstein radius
\begin{eqnarray} \label{dynamical mass}
M_{dyn}& =&  \frac{\pi}{G} \sigma_{ap}^2 R_E \left( \frac{R_E}{R_{ap}} \right)^{2-\gamma} f(\gamma)\nonumber\\
       &= &\frac{\pi}{G} \sigma_{ap}^2 D_l \theta_E \left( \frac{\theta_E}{\theta_{ap}} \right)^{2-\gamma} f(\gamma)
\end{eqnarray}
where
\begin{eqnarray} \label{f factor}
f(\gamma) &=& - \frac{1}{\sqrt{\pi}} \frac{(5-2 \gamma)(1-\gamma)}{3-\gamma} \frac{\Gamma(\gamma - 1)}{\Gamma(\gamma - 3/2)}\nonumber\\
          &\times & \left[ \frac{\Gamma(\gamma/2 - 1/2)}{\Gamma(\gamma / 2)} \right]^2
\end{eqnarray}

By combining Eq.~(\ref{lensing mass}) and Eq.~(\ref{dynamical mass}), we
obtain
\begin{equation} \label{Einstein} \theta_E =   4 \pi
\frac{\sigma_{ap}^2}{c^2} \frac{D_{ls}}{D_s} \left(
\frac{\theta_E}{\theta_{ap}} \right)^{2-\gamma} f(\gamma)
\end{equation}
Now, our observable is
\begin{equation} \label{NewObservable}
 {\cal D}^{obs} =  \frac{c^2 \theta_E }{4 \pi \sigma_{ap}^2} \left( \frac{\theta_{ap}}{\theta_E} \right)^{2-\gamma} f^{-1}(\gamma)
\end{equation}
and its theoretical counterpart (the distance ratio) ${\cal D}^{th}(z_l,z_s; {\mathbf p})$ is given by Eq.(\ref{observable}).

Fractional uncertainty of ${\cal D}$ (after calculating all relevant partial derivatives and simplifying terms) is
\begin{equation} \label{uncertainty}
\delta {\cal D} = \frac{\Delta {\cal D}}{{\cal D}} = \sqrt{4 (\delta \sigma_{ap})^2 + (1-\gamma)^2 (\delta \theta_E)^2}
\end{equation}
Following the SLACS team we took the fractional uncertainty of the
Einstein radius at the level of $5\%$ i.e. $\delta \theta_E = 0.05$ --
the same for all lenses.

Theoretically for a single system one could use $\sigma_{ap}$, but
because we deal with a sample of lenses, we shall also transform all
velocity dispersions measured within an aperture to velocity
dispersions within circular aperture of radius $R_{eff}/2$ (half the
effective radius) following the prescription of
\citet{Jorgensen95a,Jorgensen95b}: $ \sigma_0 = \sigma_{ap}
(\theta_{eff}/(2 \theta_{ap}))^{-0.04}$. In the literature it has
also been denoted as $\sigma_{e2}$. We adopt the convention to
denote observed angular Einstein radius or effective radius as
$\theta_E$ or $\theta_{eff}$ with the notation $R_E$ or $R_{eff}$
reserved for physical values (in [kpc]) of respective quantities. In
our analysis however, we will consider two cases: with $\sigma_{ap}$
and $\sigma_{0}$. While using $\sigma_{0}$, the equations
Eq.~(\ref{Einstein}), Eq.~(\ref{NewObservable}) and
Eq.~(\ref{uncertainty}) should be modified by replacing
$\sigma_{ap}$ with $\sigma_0$. In principle (i.e. within the
power-law model), the use of $\sigma_{ap}$ should work because we
rescale to the Einstein radius anyway. However, the use of
$\sigma_0$ makes our observable ${\cal D}^{obs}$ more homogeneous
for the sample of lenses located at different redshifts. One may
worry about additional uncertainties introduced into the error
budget this way since the measurement of $\theta_{eff}$ bears its
own uncertainty. Arguing that fractional uncertainty for the
effective radius is at the level of $5\%$ (like for the Einstein
radius) and bearing in mind that in the J{\o}rgensen formula
$\theta_{eff}$ is raised to a very small power one can estimate that
uncertainties of the effective radius contribute less than $1\%$ to
the uncertainty of $\sigma_0$.


Then using routines available within CosmoMC package
\citep{Lewis02}, we preformed Monte Carlo simulations of the
posterior likelihood ${\cal L} \sim \exp{(- \chi^2 / 2)}$ where
\begin{equation}
\chi^2 = \sum_{i=1}^{118} \left( \frac{{\cal D}^{th}(z_{l,i},z_{s,i}; {\mathbf p},\gamma) - {\cal D}^{obs}(\sigma_{0,i}, \theta_{E,i})}{\Delta {\cal D}^{obs}_i} \right)^2
\end{equation}
In our fits mass density power-law index $\gamma$ was taken as a free
parameter fitted together with cosmological parameters ${\mathbf
p}$. It has been suggested by \citet{Ruff11} and further supported by \citet{Brownstein12, Sonnenfeld13a}, that mass density
power-law index $\gamma$ of massive elliptical galaxies evolves with
redshift. On a combined sample of lenses from SLACS, SL2S and LSD
they fitted the $\gamma(z_l)$ data with the linear relation and
obtained $\gamma(z_l) = 2.12^{+0.03}_{-0.04} - 0.25^{+0.10}_{-0.12}
\times z_l+ 0.17^{+0.02}_{-0.02} (scatter) $. Therefore we also
performed fits assuming the linear relation: $\gamma(z_l) = \gamma_0
+ \gamma_1 z_l$ treating $\gamma_0$ and $\gamma_1$ as free
parameters together with cosmological ones.


\section{Data sets} \label{sec:data}

In order to implement the methodology described in
Section~\ref{sec:method}, we have made a comprehensive compilation
of 118 strong lensing systems from four surveys: SLACS, BELLS, LSD
and SL2S. The Sloan Lens ACS Survey (SLACS) and the BOSS
emission-line lens survey (BELLS) are spectroscopic lens surveys in
which candidates are selected respectively from Sloan Digital Sky
Survey (SDSS-III) data and Baryon Oscillation Spectroscopic Survey
(BOSS). BOSS has been initiated by upgrading SDSS-I optical
spectrographs \citep{Eisenstein11}. The idea was to take the spectra
of early type galaxies and to look for the presence of emission
lines at redshift higher than that of the target galaxy. Candidates
selected this way were followed-up with HST ACS snapshot imaging and
after image processing: subtraction of the de Vaucouleurs profile of
the target galaxy, those displaying multiple images and/or Einstein
rings have been classified as confirmed lenses. For our purpose
SLACS data comprising 57 strong lenses were taken after
\citet{Bolton08a} and \citet{Auger09}. And the BELLS data containing
25 lenses were taken from \citet{Brownstein12}.

Lenses Structure and Dynamics (LSD) survey was a predecessor of
SLACS in the sense that combined image and lens velocity dispersion
data were used to constrain the structure of lensing galaxies.
However, it was different, because lenses were selected optically
(as multiple images of sources with identified lensing galaxies) and
then followed-up spectroscopically. Therefore, in order to comply
with SLACS and BELLS, we took only 5 most reliable lenses from LSD:
CFRS03.1077; HST 14176+5226; HST 15433+5352 after \citet{Treu04},
Q0047-281 after \citet{Koopmans02}, and MG2016+112 after
\citet{Treu02}.

At last the Strong Lensing Legacy Survey (SL2S) is a project
dedicated to finding galaxy-scale lenses in the Canada France Hawaii
Telescope Legacy Survey (CFHTLS). The targets are massive red
galaxies and an automated {\tt RingFinder} software is looking for
tangentially elongated blue features around \citep{RingFinder}. If
found they are followed-up with HST and spectroscopy. The data for a
total of 31 lenses we include here were taken from
\citet{Sonnenfeld13a,Sonnenfeld13b}.

\begin{figure*}
\begin{center}
\includegraphics[angle=270,width=12cm]{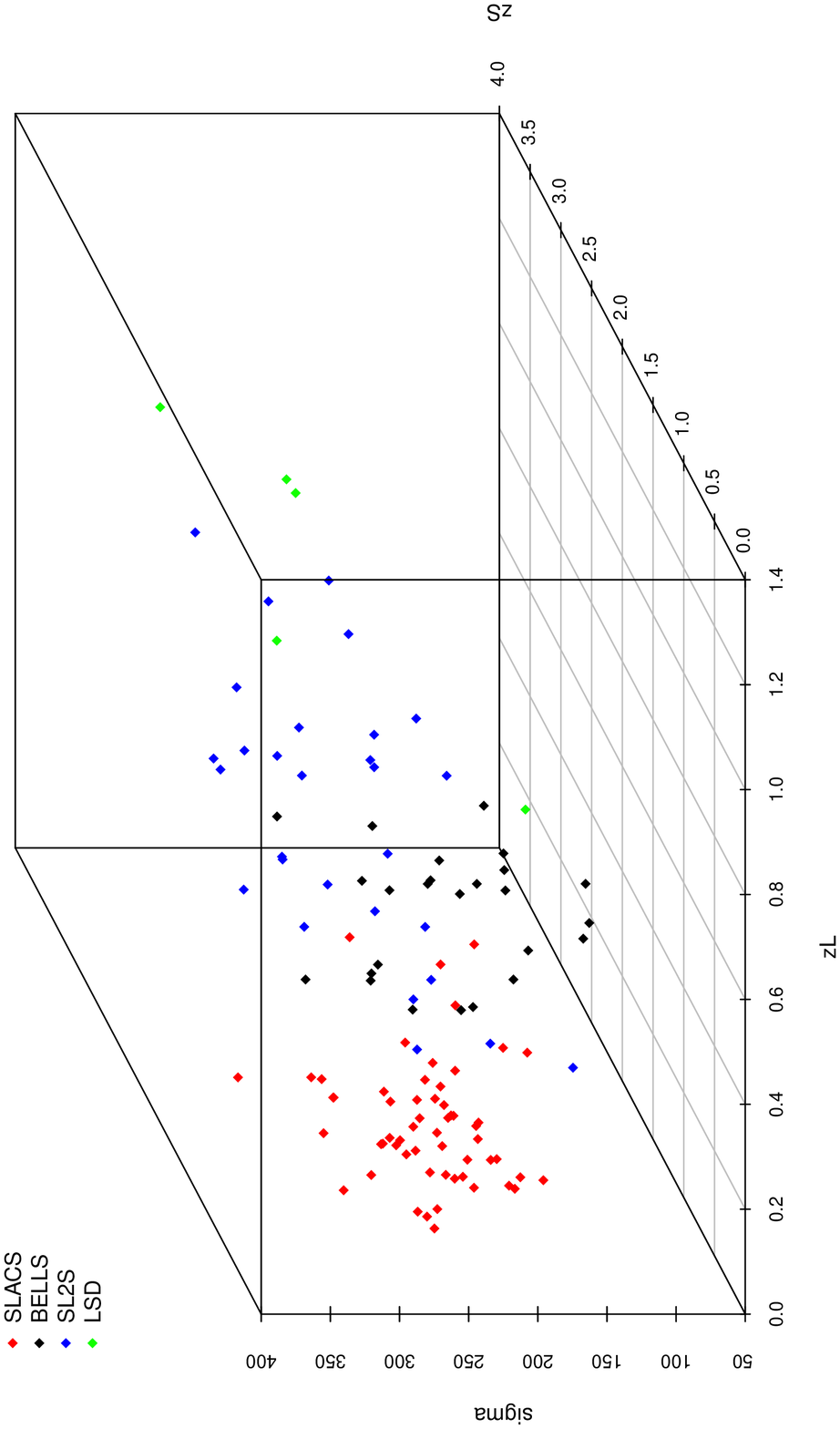}
\caption{Scatter plot of our sample of 118 strong lensing systems. One can see a fair coverage of redshifts in the combined sample.
\label{fig1}}
\end{center}
\end{figure*}

\begin{figure*}
\begin{center}
\includegraphics[angle=270,width=12cm]{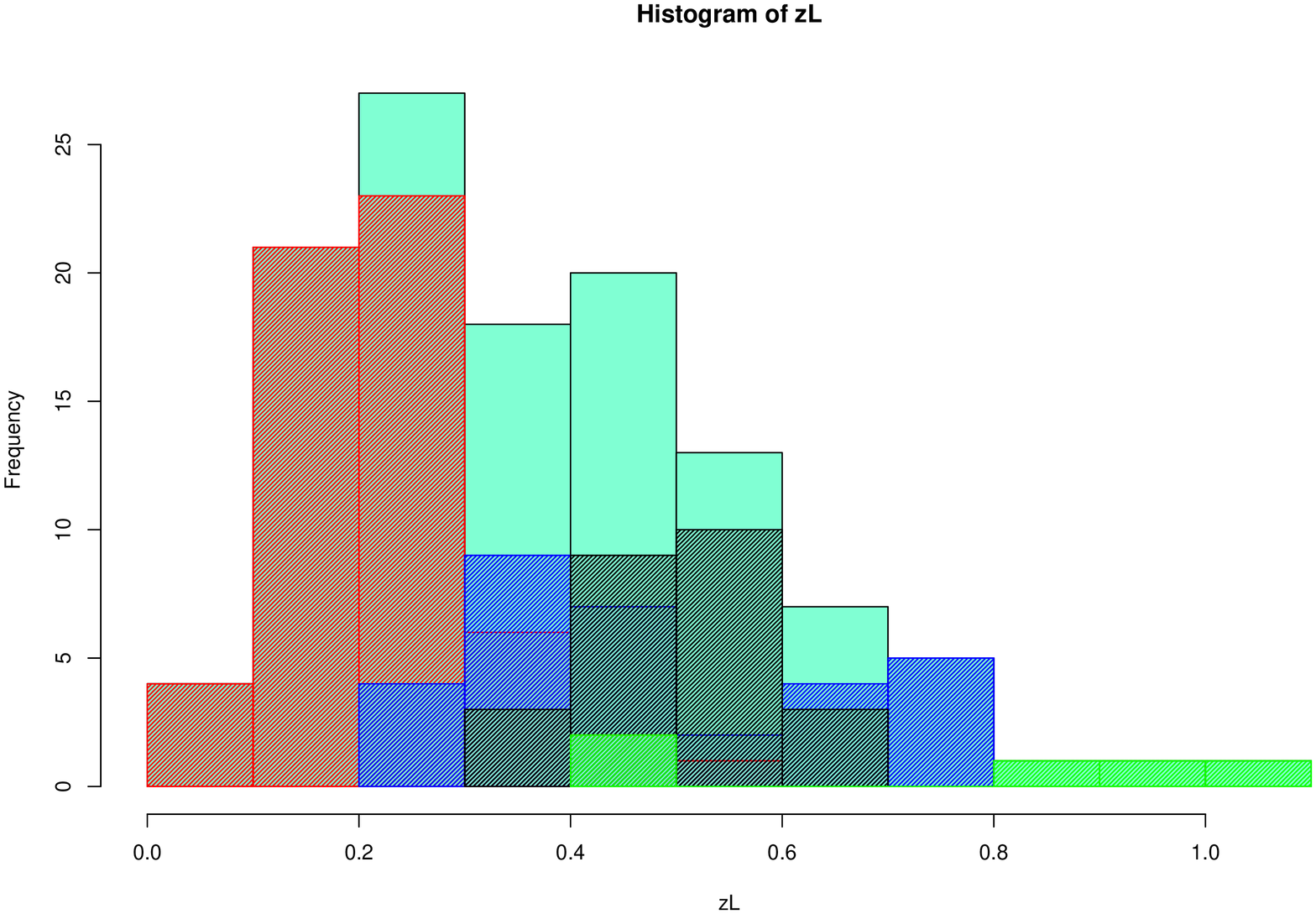}
\includegraphics[angle=270,width=12cm]{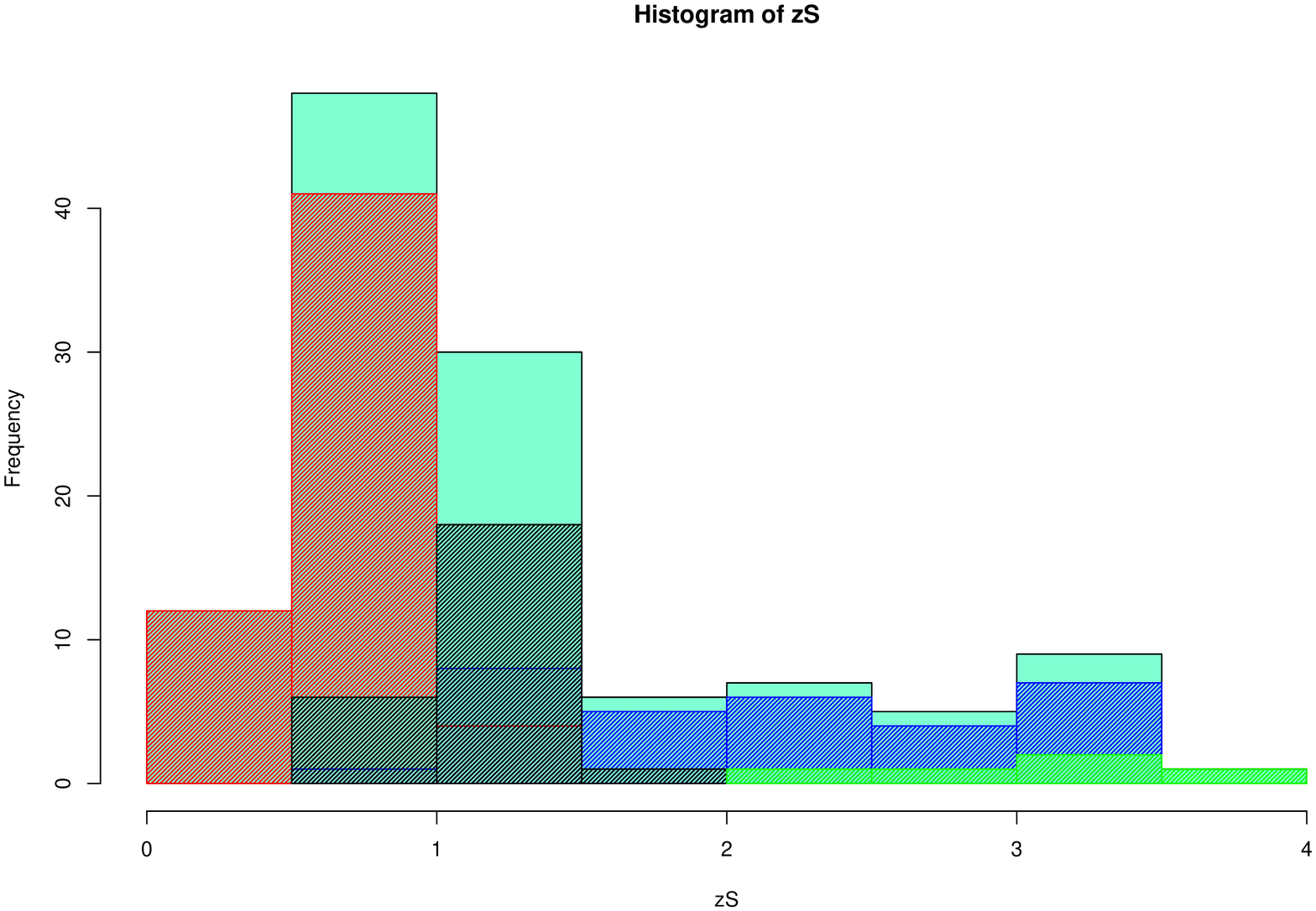}
\caption{Histogram of lens (upper panel) and source redshifts (lower panel). Different colors show
contribution of different surveys superimposed on the overall histogram. The color convention is the same as for
Fig.~\ref{fig1}. \label{fig2}}
\end{center}
\end{figure*}


Our sample is presented in Table~1. It contains all relevant data
necessary to perform cosmological model fits according to our
methodology presented in Section~\ref{sec:method}. The Einstein
radius derivation method is more or less consistent across different
surveys considered here. After subtracting de Vaucouleurs profile of
the deflector,  the Einstein radii were measured by fitting model
mass distributions to generate model lensed images and comparing
them to the observed images. The mass distribution of the main
lenses was modeled as singular isothermal ellipsoids (SIE), while in
several systems, an external shear component was also added to
describe the lensing effect of nearby groups or clusters. Even
though individual uncertainties of this procedure are different
depending on the survey and on whether the image was taken from the
Earth or from space (HST) there is a consensus that in average the
relative uncertainty of the Einstein radius is at the level of
$5\%$. Apertures for SLACS and BELLS were taken as $1.5''$ and $1''$
respectively according to the source papers. For LSD and SL2S lenses
we have assessed the aperture size from the sizes $x,y$ of the slit
reported in source papers. The last column gives the velocity
dispersion within a half effective radius calculated according to
J{\o}rgensen formula. Fig.~\ref{fig1} shows the scatter plot of our
lensing systems. One can see that inclusion of the SL2S lenses
resulted in a fair coverage of lenses and sources redshifts.

Fig.~\ref{fig2} 
presents histograms of lens and
source redshift with the contribution of respective surveys
over-plotted in the same color convention as in Fig.~\ref{fig1}.
Different surveys have the following median values of the lens redshifts:
SLACS -- $z_l = 0.215$, BELLS -- $z_l = 0.517$, LSD -- $z_l = 0.81$ and SL2S -- $ z_l = 0.456$.
The above values refer to lenses used by us. SL2S survey -- an ongoing one -- is particularly
promising for the future since it already reached the maximum lens redshift of $z_l = 0.8$.

\section{Results} \label{sec:results}

\begin{figure*}
\begin{center}
\includegraphics[width=8cm]{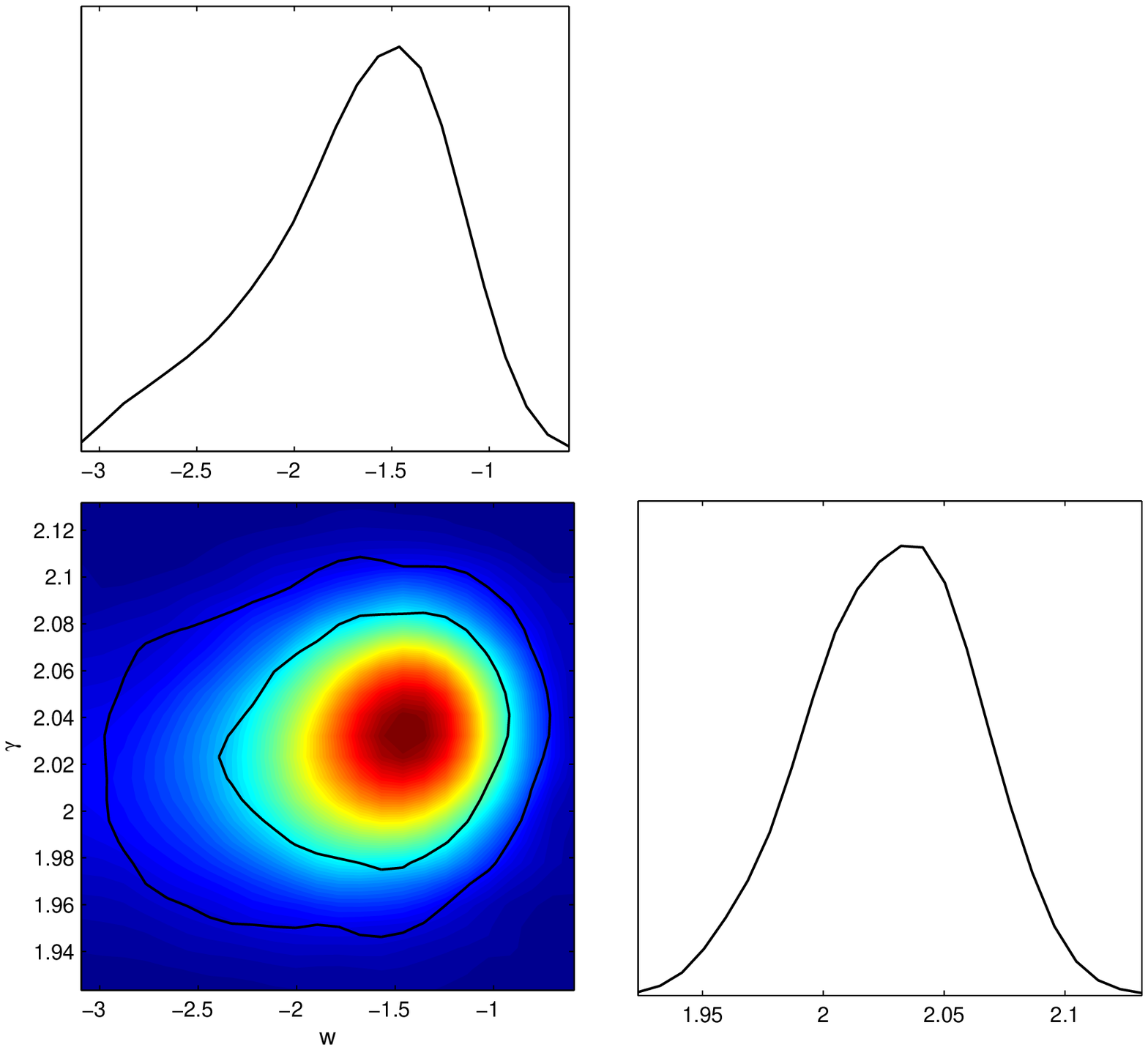}\includegraphics[width=8cm]{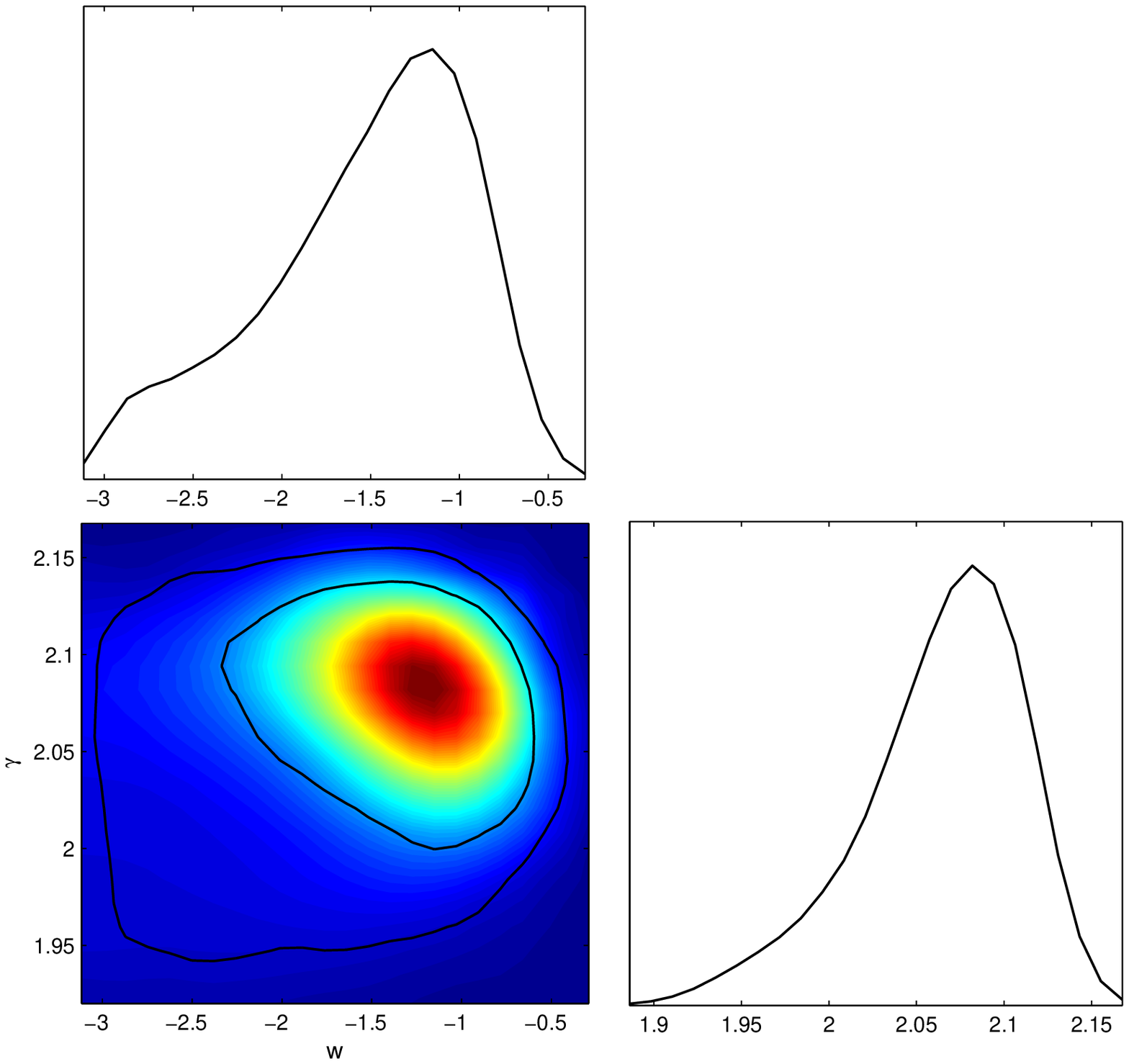}
\caption{Joint fits of mass density slope $\gamma$ and $w$
coefficient in the XCDM model. Left panel shows the results obtained
with velocity dispersion within the aperture $\sigma_{ap}$ while on
the right panel corrected velocity dispersion $\sigma_0$ was used.
$\Omega_m=0.315$ is assumed based on the Planck observations
\citep{Ade14}. Marginalized probability density functions for
$\gamma$ and $w$ are also shown. \label{fig3}}
\end{center}
\end{figure*}

\begin{figure*}
\begin{center}
\includegraphics[width=8cm]{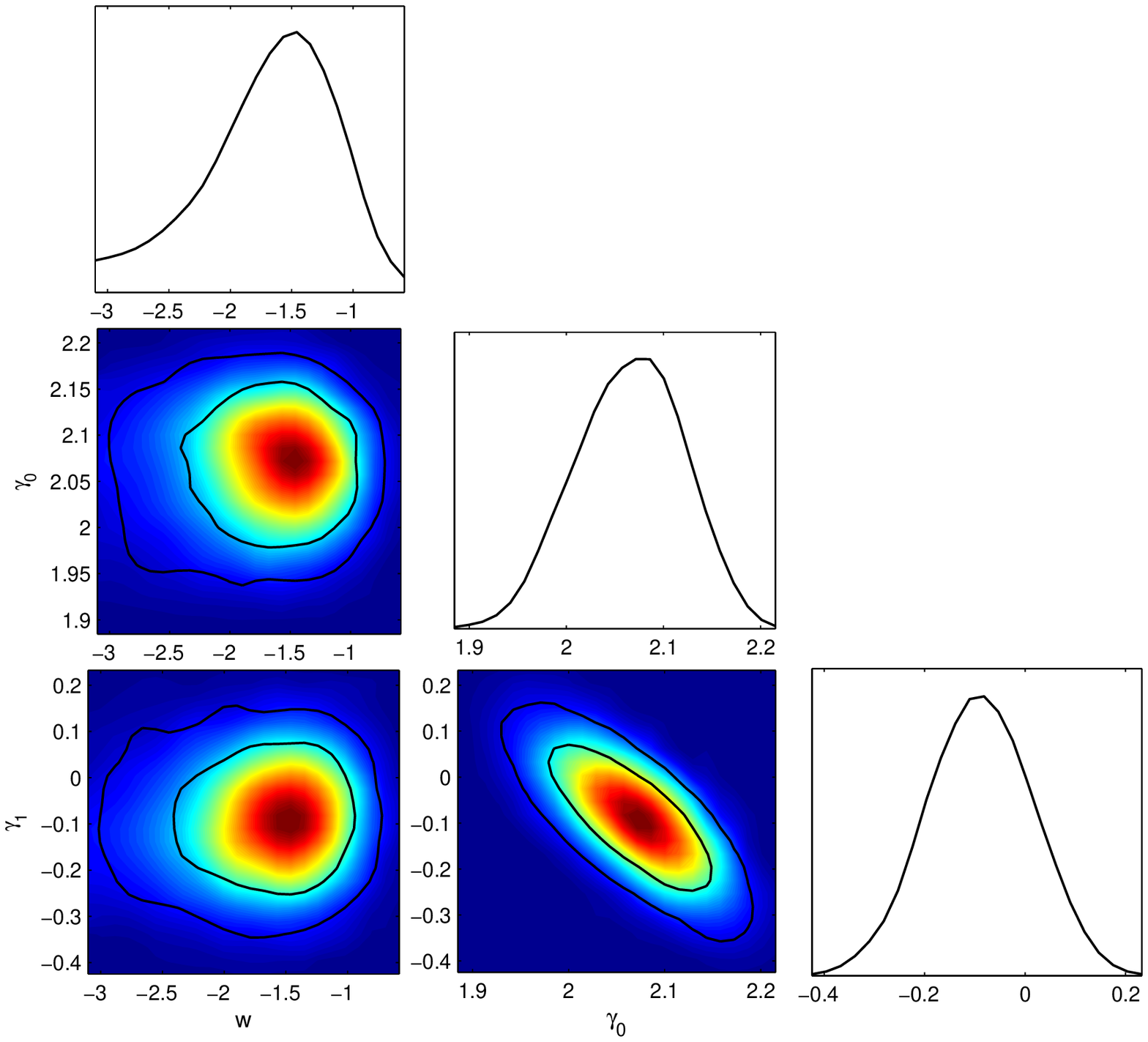}\includegraphics[width=8cm]{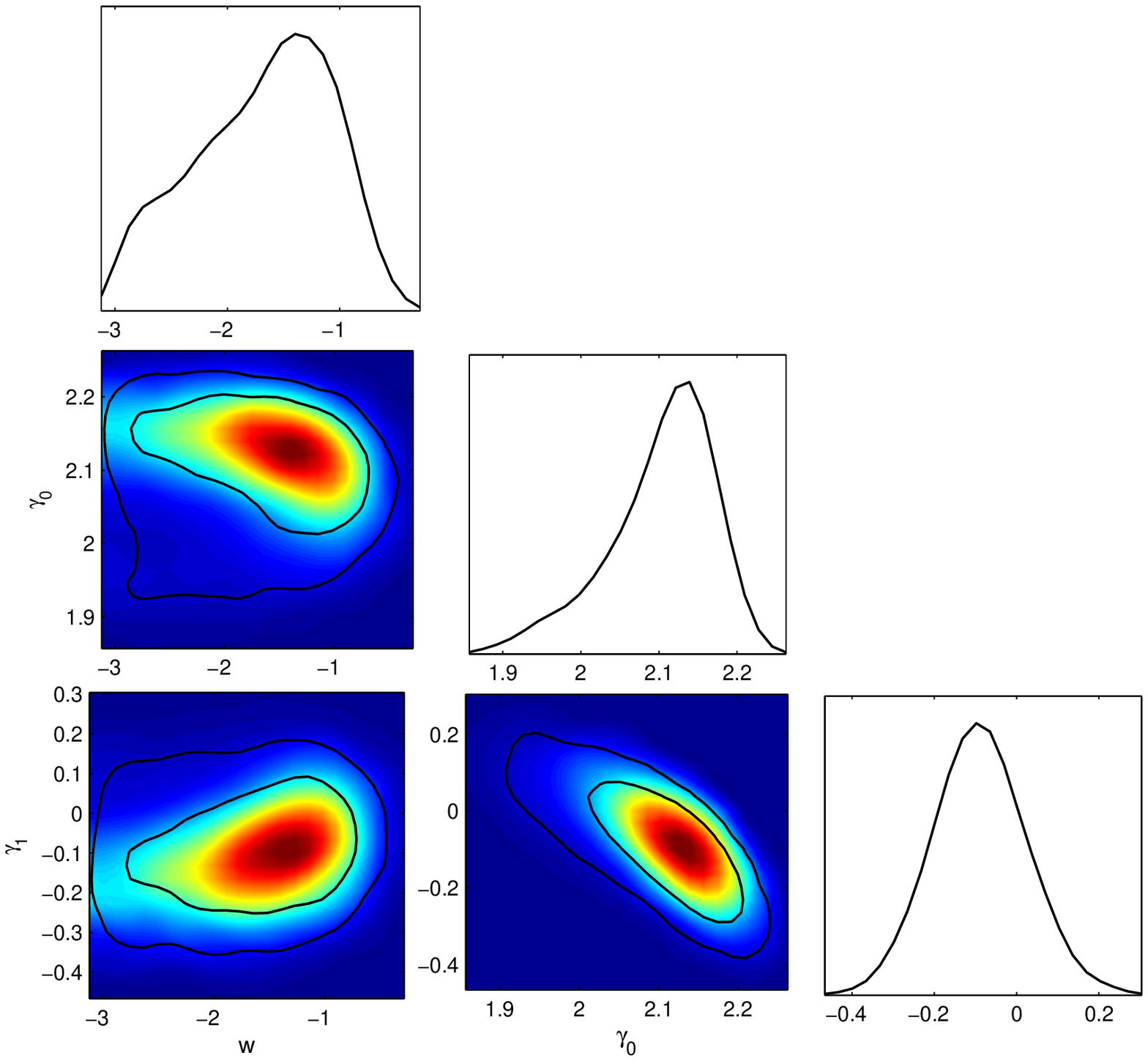}
\caption{Joint fits of $w$ coefficient in the XCDM model and mass
density slope parameters ($\gamma_0$, $\gamma_1$) in evolving slope
scenario $\gamma(z) = \gamma_0 + \gamma_1 z_l$. Left and right
panels display the results obtained by using $\sigma_{ap}$ and
$\sigma_0$ respectively. \label{fig4}}
\end{center}
\end{figure*}

\begin{figure*}
\begin{center}
\includegraphics[width=8cm]{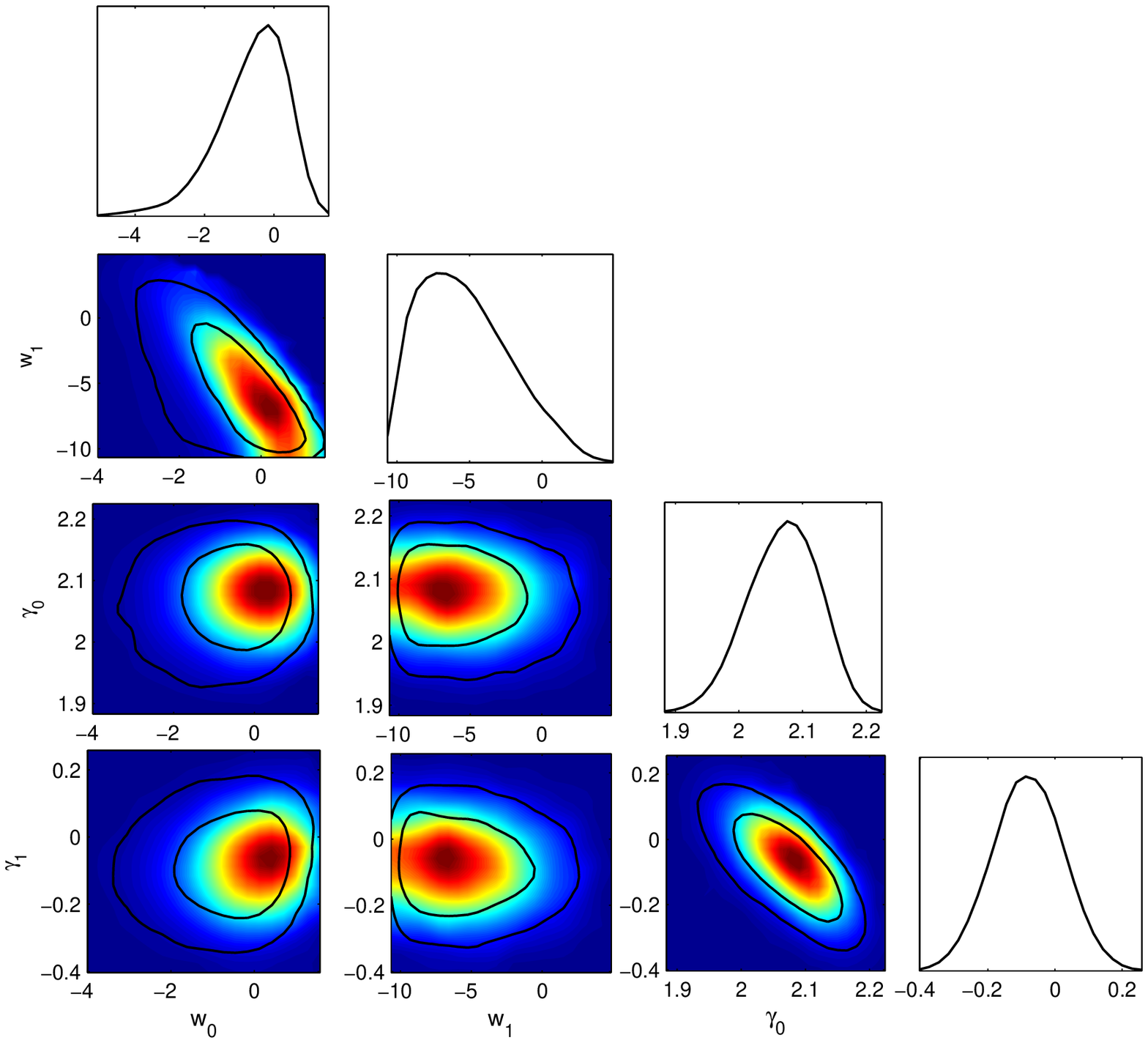}\includegraphics[width=8cm]{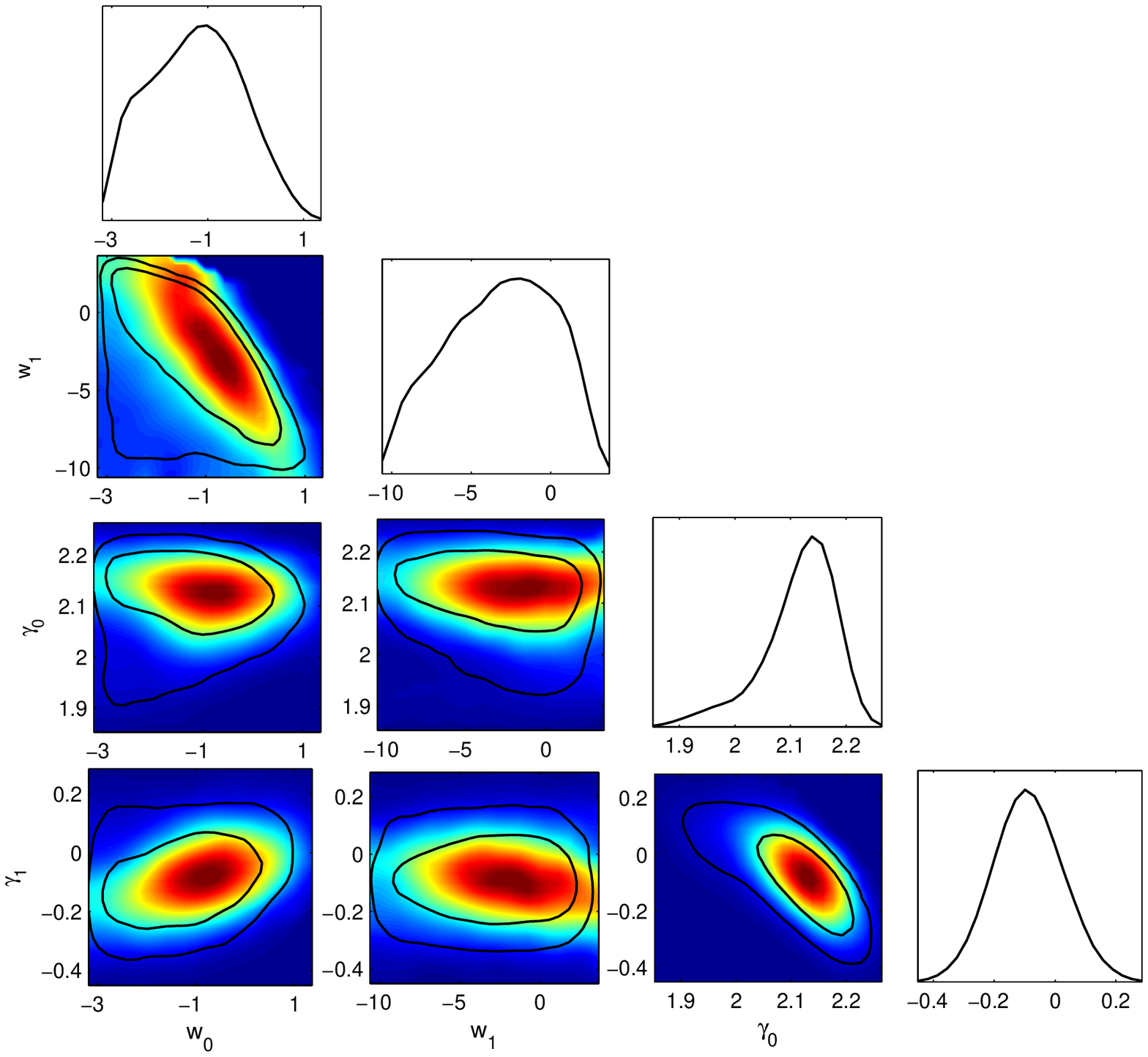}
\caption{ Joint fits of $(w_0, w_1)$ evolving cosmic equation of
state coefficients in the CPL parametrization. Lensing galaxies were
assumed to have evolving mass density slope $\gamma(z) = \gamma_0 +
\gamma_1 z_l$ and $(\gamma_0, \gamma_1)$ parameters. Left and right
panels display the results obtained by using $\sigma_{ap}$ and
$\sigma_0$ respectively.
 \label{fig5}}

\end{center}
\end{figure*}

\begin{figure*}
\begin{center}
\includegraphics[width=8cm]{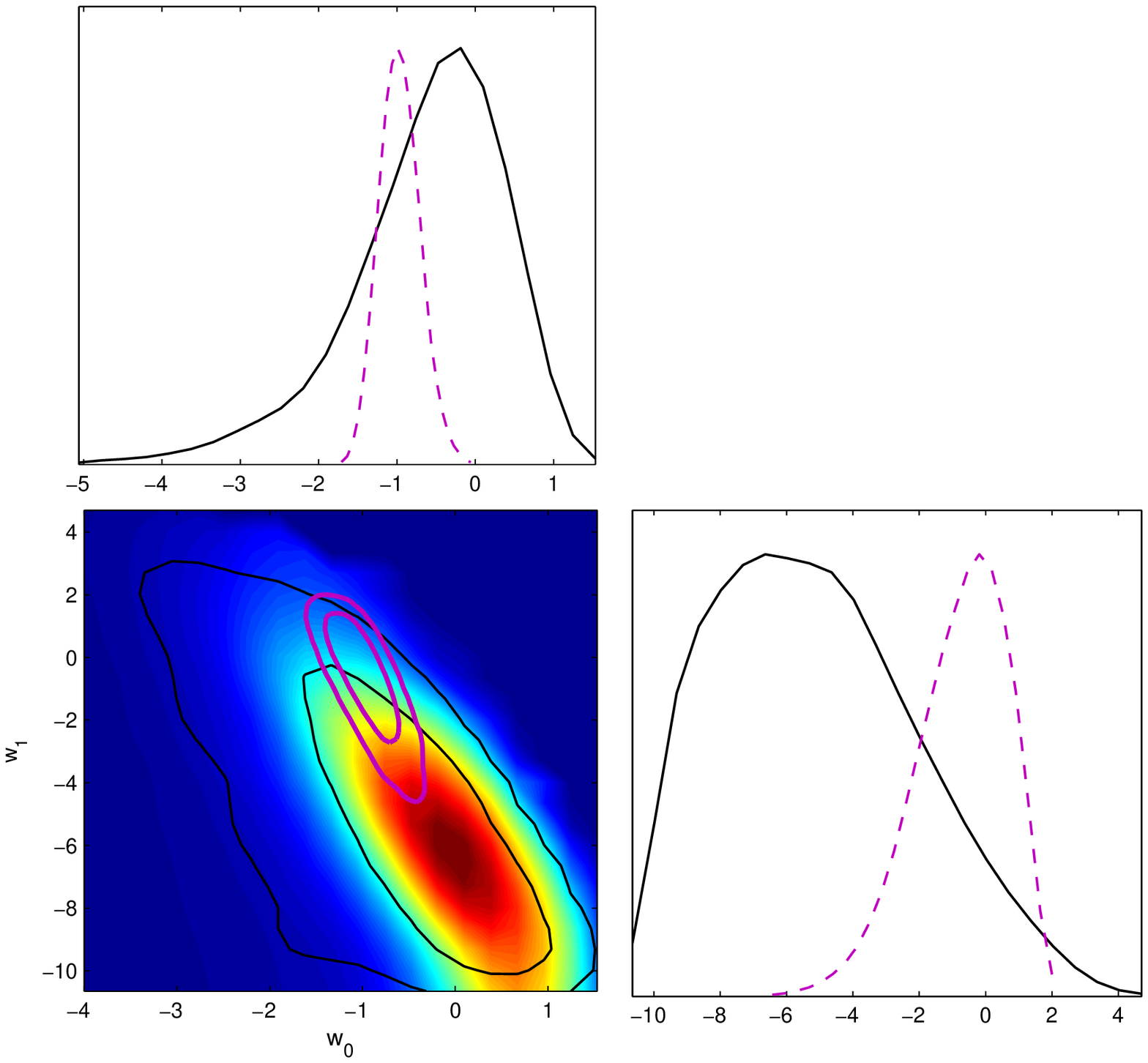}\includegraphics[width=8cm]{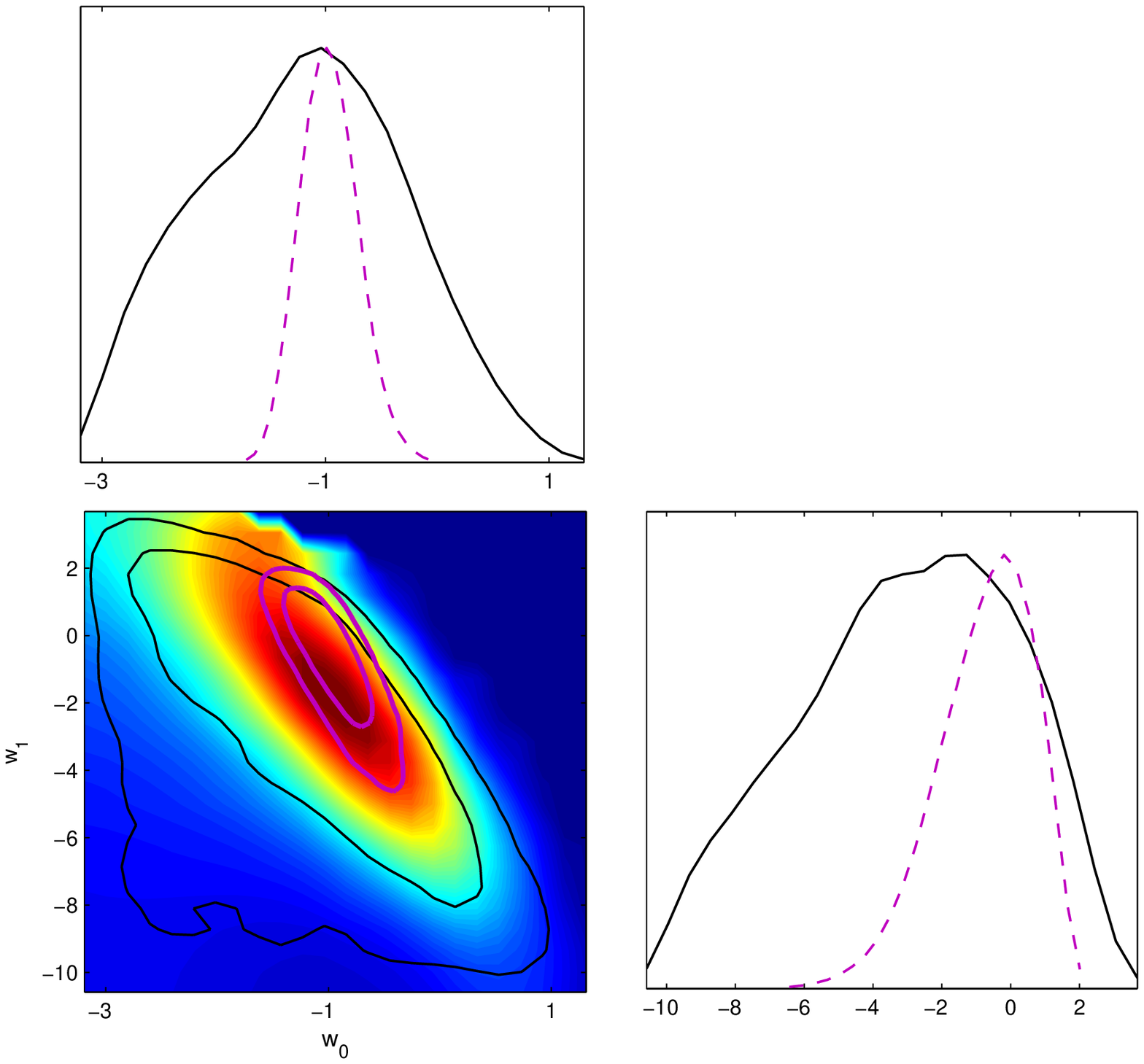}
\caption{Joint fits of $(w_0, w_1)$ evolving cosmic equation of
state coefficients in the CPL parametrization. Lensing galaxies were
assumed to have evolving mass density slope $\gamma(z) = \gamma_0 +
\gamma_1 z_l$ and $(\gamma_0, \gamma_1)$ parameters were fixed at
our best-fit values in $XCDM$ model. Left and right panels display
the results obtained by using $\sigma_{ap}$ and $\sigma_0$
respectively.
 \label{fig6}}

\end{center}
\end{figure*}

Although the lens redshift $z_l$ and source redshift $z_ s$ both
cover a wide range in our sample, distance ratio $\mathcal{D}$ is
still confined to a very compact range of values, which leads to
poor constraining power for $\Omega_m$ parameter \citep{Biesiada10}.
Therefore in this paper, we fix $\Omega_m$ at the best-fit value
$\Omega_m=0.315$ based on the recent Planck observations
\citep{Ade14}. This disadvantage of our method, i.e. the necessity
of taking a prior for the matter density parameter, is to a certain
degree alleviated by the benefit of being independent of the Hubble
constant which cancels in the distance ratio. Consequently $H_0$ and
its uncertainty do not influence the results. Performing fits to
different cosmological scenarios on the $n=118$ strong lensing
sample, we obtain the results displayed in Table~2. The marginalized
probability distribution of each parameter and the marginalized 2D
confidence contours are presented in Fig.~\ref{fig3}-\ref{fig6}.

We started our analysis with dark energy phenomenon modeled by
barotropic fluid having constant equation of state coefficient $w$
and we considered two cases of mass density profiles in lenses: a
non-evolving power-law density profile and an evolving one (denoted
in Table~2 as $XCDM1$ and $XCDM2$, respectively). Power-law
exponents $\gamma$, $\gamma_0$ and $\gamma_1$ were treated as free
parameters to be fitted.

First, we consider fits on our sample taking the aperture velocity
dispersion $\sigma_{ap}$ as the lens parameter. In the $XCDM$ model
where $w$ is the only free cosmological parameter and $\gamma$ is
the mass density power-law index parameter, we obtain
$w=-1.45^{+0.54}_{-0.95}$ and $\gamma=2.03\pm 0.06$. In the second
case, when the power-law mass density profile was allowed to evolve:
$\gamma(z_l) = \gamma_0 + \gamma_1 z_l$, the best-fit values for the
parameters are $w = -1.48^{+0.54}_{-0.94}$, $\gamma_0=2.06\pm 0.09$,
and $\gamma_1=-0.09\pm0.16$. One can see that $w$ coefficient
obtained from the strong lensing sample is consistent with the
$\Lambda$CDM model ($w=-1$) at $1\sigma$ level. Fits on the $\gamma$
parameter also reveal compatibility between our sample of lenses and
the previous smaller combined sample from SLACS, SL2S and LSD
\citep{Ruff11,Sonnenfeld13b}. In an attempt to constrain cosmology
with the CPL parametrization describing an evolving cosmic equation
of state, we first consider the case that both $w_0$, $w_1$ and
$\gamma_0$, $\gamma_1$ are free parameters (denoted in Table~2 as
$CPL1$). By fitting the CPL model to the full $n=118$ sample with
$\sigma_{ap}$, we get the marginalized $1\sigma$ constraints of the
parameters $w_0=-0.15^{+1.27}_{-1.60}$, $w_1=-6.95^{+7.25}_{-3.05}$
and $\gamma_0=2.08\pm 0.09$, $\gamma_1=-0.09\pm 0.17$. By fixing
$\gamma$ parameters at our best-fit values (denoted in Table~2 as
$CPL2$), the best-fit values for the two cosmic equation of state
parameters are: $w_0=-0.16^{+1.21}_{-1.48}$ and
$w_1=-6.25^{+6.25}_{-3.75}$. We also show the marginalized $1\sigma$
and $2\sigma$ contours of the two parameter in Fig.~\ref{fig6}. It
can be seen that, comparing to the previous analysis with a smaller
sample \citep{Biesiada10,Cao12}, fits for $w_0$ and $w_1$ are
significantly improved with a larger strong lensing sample.

Working on the sample with the aperture corrected velocity
dispersion $\sigma_{0}$, we find that the dark energy equation of
state parameter ($w=-1.15^{+0.56}_{-1.20}$ and
$w=-1.35^{+0.67}_{-1.50}$) agrees very well with the respective
value derived from Planck observations combined with BAO data
($w=-1.13^{+0.13}_{-0.10}$) \citep{Ade14}. Constraints on the mass
density power-law index parameters ($\gamma_0=2.13^{+0.07}_{-0.12}$,
$\gamma_1=-0.09\pm0.17$) are even more consistent with the previous
analysis \citep{Ruff11, Sonnenfeld13b} especially with $\sigma_0$
taken as the lens parameter. In the case of evolving cosmic equation
of state (CPL parametrization), we obtain
$w_0=-1.00^{+1.54}_{-1.95}$, $w_1=-1.85^{+4.85}_{-6.75}$ and
$\gamma_0=2.14^{+0.07}_{-0.10}$, $\gamma_1=-0.10\pm 0.18$. By fixing
$\gamma$ parameters at the best-fit values, the fits for $w_0$ and
$w_1$ are: $w_0=-1.05^{+1.43}_{-1.77}$ and
$w_1=-1.65^{+4.25}_{-6.35}$. One can see fairly good agreement
between our results obtained by using distance ratio method for
strong lensing systems and the concordance $\Lambda$CDM model. In
order to compare our fits with the results obtained using supernovae
Ia, likelihood contours obtained with the latest Union2.1
compilation \citep{Suzuki12} consisting of 580 SN Ia data points are
also plotted in Fig.~6. For a fair comparison, the matter density
parameter $\Omega_m$ is also fixed at the Planck best-fit value
$\Omega_m=0.315$ and the systematic errors of observed distance
moduli are also considered in the likelihood calculation
\citep{Cao14}. We see that 1$\sigma$ confidence regions from the two
data sets overlap very well with each other. This means that the
results obtained on the sample of strong lenses are consistent with
the SNIa fits. This consistency at 1$\sigma$ level is different from
earlier results obtained with smaller sample of lenses
\citep{Biesiada10,Cao12}.

One can clearly see from Fig~\ref{fig6} that principal axes of
confidence regions obtained with supernovae and strong lenses are
inclined at higher angles that in previous studies
\citep{Biesiada10,Biesiada11,Cao12}. This sustains the hope that
careful choice of the sample in terms of lens and source redshifts,
would eventually realize an ultimate dream to have a complementary
probe breaking the degeneracy in $(w_0,w_1)$ plane (discussed in
\citet{Linder04,Piorkowska13}). This will be a subject of another
study.

Our method based on distance ratio for strong gravitational lensing
systems may also contribute to testing the consistency between
luminosity and angular diameter distances \citep{Cao11a,Cao11b}. As
it is well known, these two observables are related to each other
via Etherington reciprocity relation. Earlier discussions of this
issue can be found in \citet{Bassett04,Uzan04,HLB10}.

\begin{figure}
\begin{center}
\includegraphics[angle=270,width=70mm]{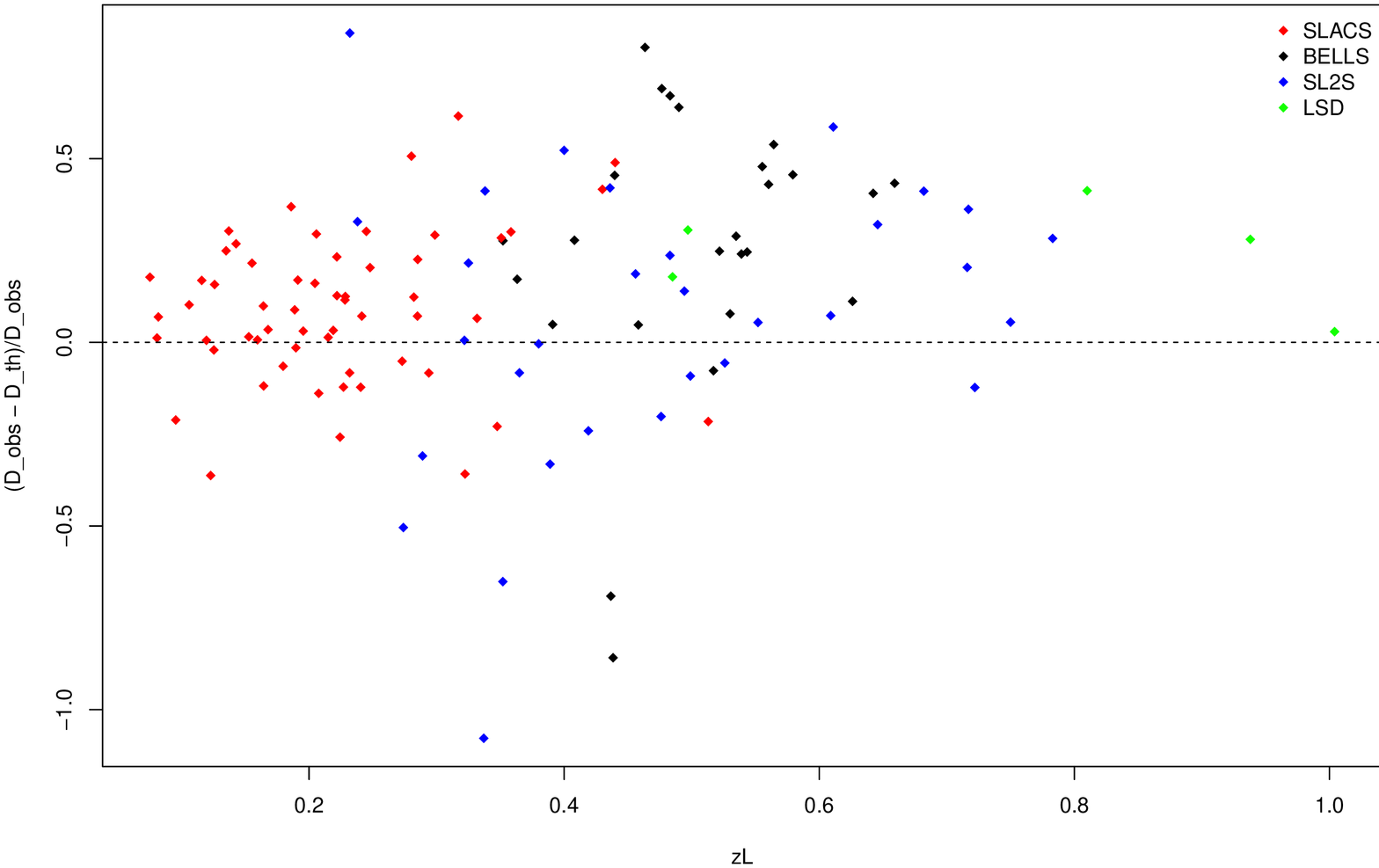}
\includegraphics[angle=270,width=70mm]{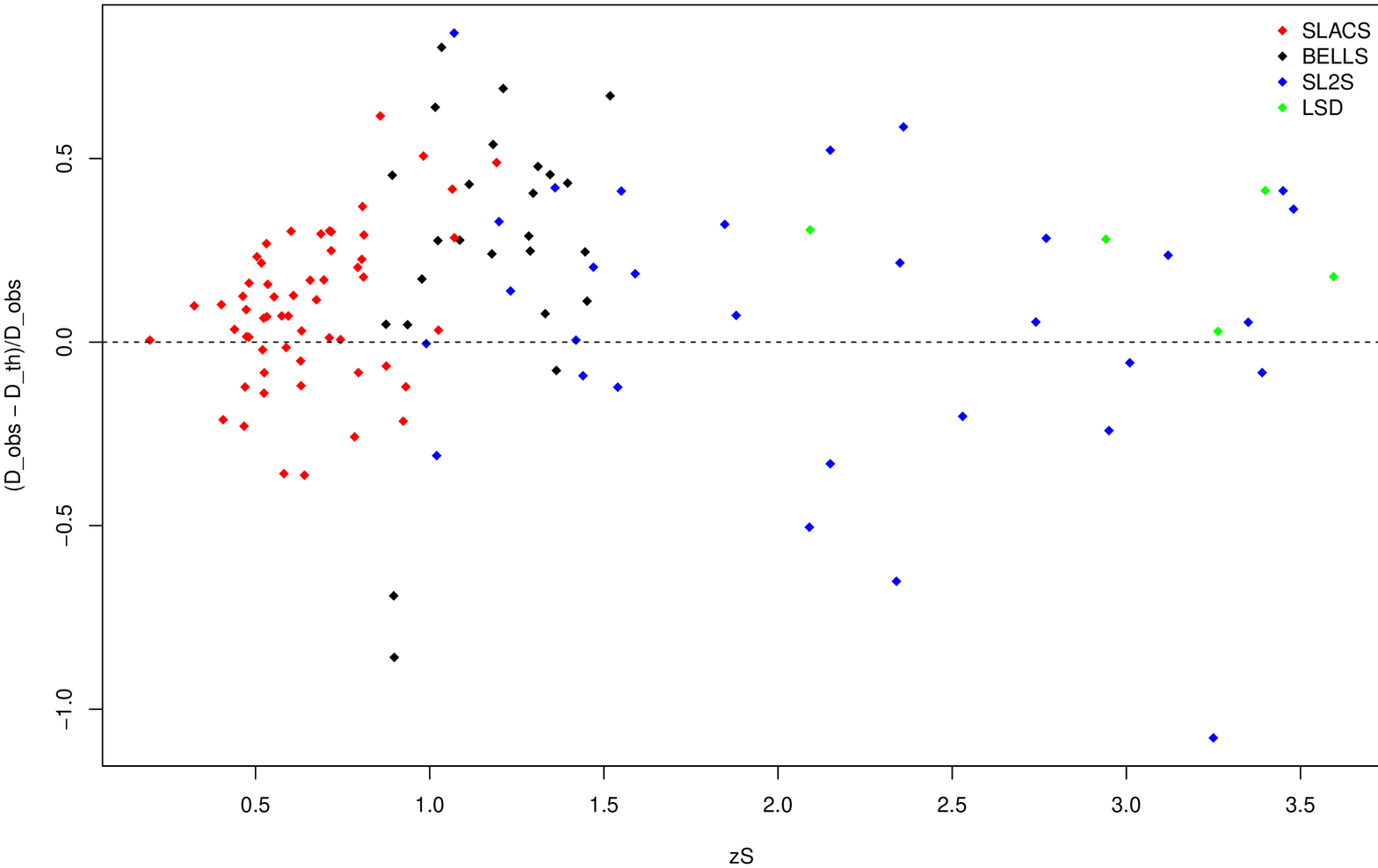}
\includegraphics[angle=270,width=70mm]{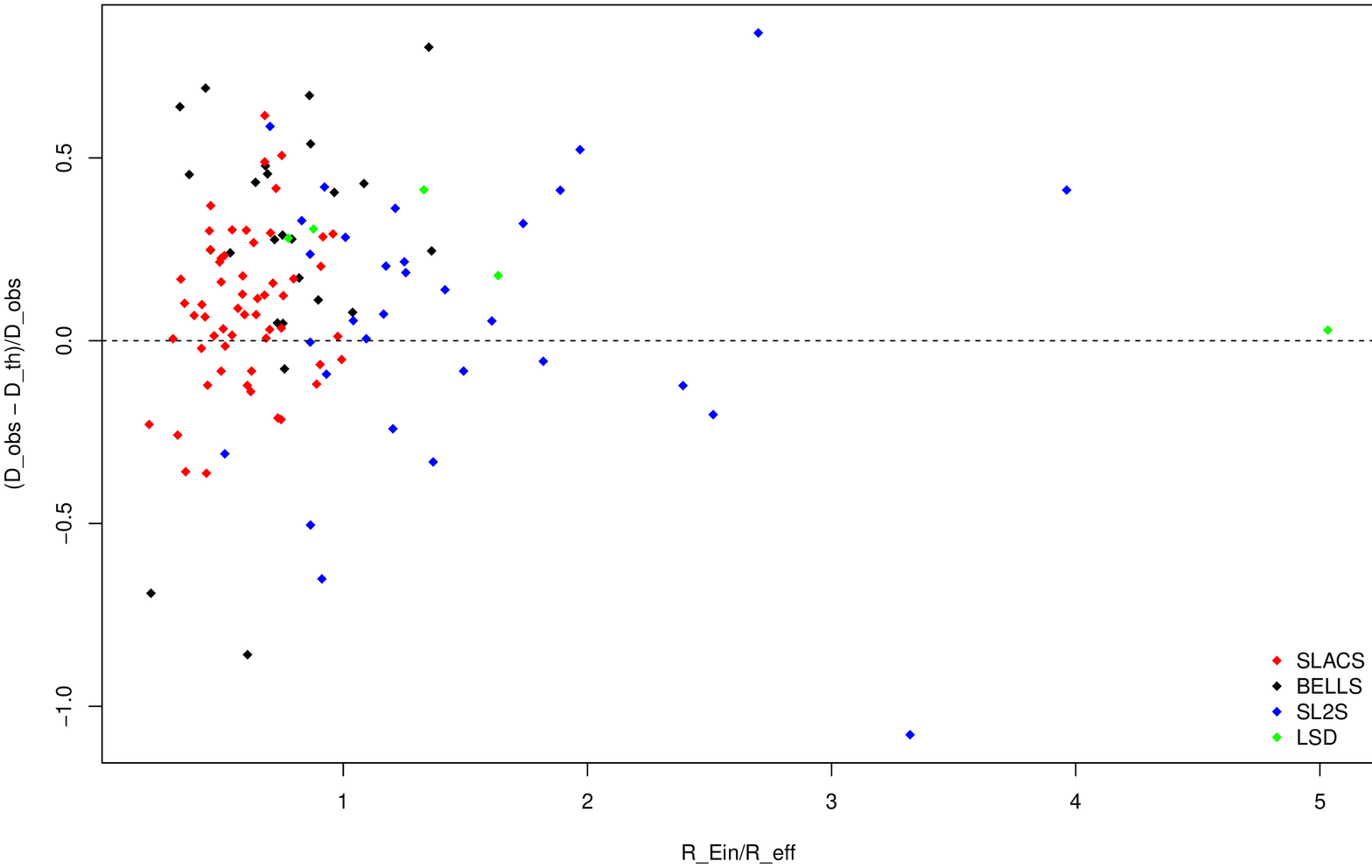}
\includegraphics[angle=270,width=70mm]{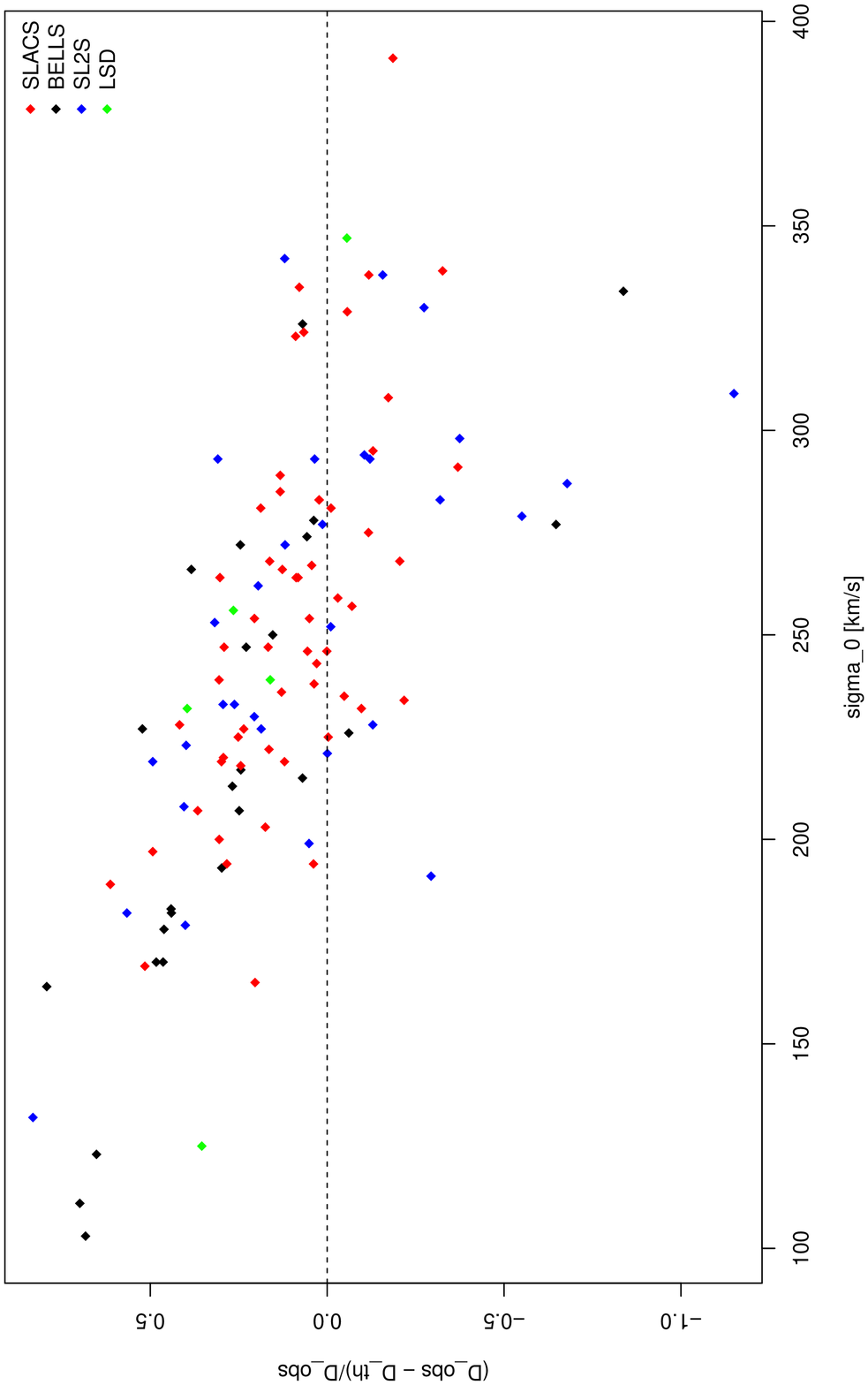}
\end{center}
\caption{Relative residuals in our observable: $({\cal D}^{obs} -
{\cal D}^{th})/{\cal D}^{obs}$ as a function of lens redshift $z_l$,
source redshift $z_s$, Einstein radius relative to the effective
radius $R_E/R_{eff}$ and the aperture corrected velocity dispersion
$\sigma_0$. \label{residuals}}
\end{figure}

In order to study the systematics and scatter in our method we
performed the diagnostics of residuals. Plots of relative residuals
$({\cal D}^{obs} - {\cal D}^{th})/{\cal D}^{obs}$ as a function of
$z_l$, $z_s$, $R_E/R_{eff}$ and $\sigma_0$ for the case of $XCDM$
cosmology, assuming non-evolving mass density profile in lenses, are
displayed in Fig.~\ref{residuals}. This figure is representative of
similar diagnostics for the CPL case and evolving mass density
profile. One can see that there is no correlation between residuals
and the redshifts of lenses or sources, as well as with the Einstein
radius relative to the effective radius. However, there is a
noticeable anti-correlation (correlation coefficient ca. -0.6) with
the velocity dispersion. This trend is especially pronounced for
lenses with $\sigma_0 < 230 \; km/s$. If one excluded small velocity
dispersion lenses the result would be comparable to other
diagnostics discussed. This effect is probably related to recent
finings of \citet{Shu14} who extended SLACS survey into lower mass
region and found that elliptical galaxies with smaller observed
velocity dispersions are more centrally concentrated. In any case,
however there is a considerable scatter left (at the level of $\pm
50\%$). This scatter could be attributable to individual properties
of lenses, like their environment or deviation form the spherical
symmetry. From theoretical point of view it has been known for a
long time \citep{Falco85} that there exists a transformation of mass
distribution in the lens that leaves all observables invariant. This
is so called ``mass-sheet'' degeneracy --- a topic which has
recently revived in the context of double Einstein Ring lenses
\citep{SchneiderSluse}. However, from the physical point of view one
of the main ingredients of the ``mass-sheet'' degeneracy comes from
the contamination of secondary lenses (clumps of matter along or
close to the line of sight). Such factors have not been considered
in our methodology and their proper inclusion is still challenging.

\section{Conclusions} \label{sec:conclusions}

In conclusion, our analysis demonstrates that strong gravitationally
lensed systems can already now be used to probe cosmological
parameters, especially the cosmic equation of state for dark energy.
One may say that the approach initiated in
\citet{Biesiada06,Grillo08,Biesiada10,Cao12} can be further
developed.

However, there are several sources of systematics we do not consider
in this paper and which remain to be addressed in the future
analysis. Let us start with simplified assumptions underlying our
method. The first one is related to the interpretation of observed
velocity dispersions. In this paper, we adopted the spherical Jeans
equation to connect observed velocity dispersions to the masses and
we did it assuming that anisotropy $\beta$ parameter was zero. As
shown in \citet{Koopmans06} and in more detailed way in
\citet{Koopmans05} (the equations generalizing our
Eq.~\ref{dynamical mass}-\ref{Einstein} to arbitrary $\beta$ can be
found there) anisotropy parameter is degenerated with the slope
$\gamma$. Therefore in our approach power-law index should
understood as an effective descriptor capturing both the density
profile and anisotropy of the velocity dispersions. This effect
clearly contributes to the scatter in our results. Another issue is
the three-dimensional shape of lensing galaxies, the
prolateness/oblateness of lensing galaxies can systematically bias
the connection between the mass and the velocity dispersion
\citep{Chae03}. This effect also contributes to the scatter at high
redshifts and might reveal as a systematic effect at low redshifts.
All these effects are hard to be rigorously accounted in context of
cosmological studies like in this paper. Recent results on this
issue can be found in \citet{Barnabe14}.

The other systematic is the influence of the line of sight
(foreground and background) contamination. The problem has been
recognized long time ago \citep{Kochanek88,Bar-Kana96,Keeton97} with
a heuristic suggestion that adding an external shear to an elliptic
lens model greatly improves the fits of multiple image
configurations. \citet{Wambsganss05} addressed the question of
secondary lenses on the line of sight and concluded that the role of
secondary lenses is a strong function of source redshift and can be
important in 38\% of cases for a source at $z_s=3.5$. High redshift
sources are advantageous from the point of view of dark energy
studies and at the same time they are challenging from the point of
view of line of sight contamination. One of the most recent studies
\citep{Jaroszyki12} trying to quantify the influence of the matter
along line of sight on strong lensing used the technique of
simulations of many multiple image configurations using a realistic
model of light propagation in an inhomogeneous Universe model (based
on the Millenium simulation). Further progress in this direction has
recently been achieved by \citet{panglos} in a paper accompanied
with publicly available code {\tt Pangloss}. They used a simple halo
model prescription for reconstructing the mass along a line of sight
up to intermediate redshifts and calibrated their procedure with
ray-tracing through the Millenium Simulation.

Finally, as proposed by \citet{Rusin03}, another approach to
constrain the power-law index $\gamma$ of lensing galaxies is to
assume a self-similar mass profile and combine different strong lens
systems to statistically constrain the mass profile, which could be
combined with the method considered in this paper to break the
degeneracy between $\gamma$ and cosmological parameters. Inclusion
of this combination into cosmological use of strong lensing systems
will be a subject of a separate paper.

\section*{Acknowledgements}

The authors are grateful to the referee for very useful comments which allowed to improve the paper.
This work was supported by the Ministry of Science and Technology
National Basic Science Program (Project 973) under Grants Nos.
2012CB821804 and 2014CB845806, the Strategic Priority Research
Program ``The Emergence of Cosmological Structure" of the Chinese
Academy of Sciences (No. XDB09000000), the National Natural Science
Foundation of China under Grants Nos. 11373014 and 11073005, the
Fundamental Research Funds for the Central Universities and
Scientific Research Foundation of Beijing Normal University, and
China Postdoctoral Science Foundation under grant No. 2014M550642.
R.G. acknowledges support from the CNES.

Part of the research was conducted within the scope of the HECOLS International Associated Laboratory,
supported in part by the Polish NCN grant DEC-2013/08/M/ST9/00664 - M.B., A.P., R.G. gratefully acknowledge this support.
M.B. obtained approval of foreign talent introducing project in
China and gained special fund support of foreign knowledge
introducing project. He also gratefully acknowledges hospitality of
Beijing Normal University.

\begin{deluxetable}{lcccccccc} \label{data}
\tablewidth{0pt} \tablecaption{Compilation of strong lensing
systems} \tablehead{ \colhead{Name}           & \colhead{$z_l$} &
\colhead{$z_s$}          & \colhead{$\sigma_{ap}\; [km/s]$}  &
\colhead{$\theta_E \;['']$}     & \colhead{survey}    & \colhead{$\theta_{ap}\;['']$}
& \colhead{$\theta_{eff}\; ['']$}  & \colhead{$\sigma_0 \; [km/s]$} } \startdata

J0151+0049 &    0.517 & 1.364 & 219$\pm$39 &    0.68 &  BELLS & 1 & 0.89 &  226$\pm$40 \\
J0747+5055 &    0.438 & 0.898 & 328$\pm$60 &    0.75 &  BELLS & 1 & 1.24 &  334$\pm$61 \\
J0747+4448 &    0.437 & 0.897 & 281$\pm$52 &    0.61 &  BELLS & 1 & 2.87 &  277$\pm$51 \\
J0801+4727 &    0.483 & 1.518 & 98$\pm$24 & 0.49 &  BELLS & 1 & 0.57 &  103$\pm$25 \\
J0830+5116 &    0.53 &  1.332 & 268$\pm$36 &    1.14 &  BELLS & 1 & 1.1 &   274$\pm$37 \\
J0944-0147 &    0.539 & 1.179 & 204$\pm$34 &    0.72 &  BELLS & 1 & 1.35 &  207$\pm$35 \\
J1159-0007 &    0.579 & 1.346 & 165$\pm$41 &    0.68 &  BELLS & 1 & 0.99 &  170$\pm$42  \\
J1215+0047 &    0.642 & 1.297 & 262$\pm$45 &    1.37 &  BELLS & 1 & 1.42 &  266$\pm$46 \\
J1221+3806 &    0.535 & 1.284 & 187$\pm$48 &    0.7 &   BELLS & 1 & 0.93 &  193$\pm$49 \\
J1234-0241 &    0.49 &  1.016 & 122$\pm$31 &    0.53 &  BELLS & 1 & 1.61 &  123$\pm$31 \\
J1318-0104 &    0.659 & 1.396 & 177$\pm$27 &    0.68 &  BELLS & 1 & 1.06 &  182$\pm$28 \\
J1337+3620 &    0.564 & 1.182 & 225$\pm$35 &    1.39 &  BELLS & 1 & 1.6 &   227$\pm$35 \\
J1349+3612 &    0.44 &  0.893 & 178$\pm$18 &    0.75 &  BELLS & 1 & 2.03 &  178$\pm$18 \\
J1352+3216 &    0.463 & 1.034 & 161$\pm$21 &    1.82 &  BELLS & 1 & 1.35 &  164$\pm$21 \\
J1522+2910 &    0.555 & 1.311 & 166$\pm$27 &    0.74 &  BELLS & 1 & 1.08 &  170$\pm$28 \\
J1541+1812 &    0.56 &  1.113 & 174$\pm$24 &    0.64 &  BELLS & 1 & 0.59 &  183$\pm$25 \\
J1542+1629 &    0.352 & 1.023 & 210$\pm$16 &    1.04 &  BELLS & 1 & 1.45 &  213$\pm$16 \\
J1545+2748 &    0.522 & 1.289 & 250$\pm$37 &    1.21 &  BELLS & 1 & 2.65 &  247$\pm$37 \\
J1601+2138 &    0.544 & 1.446 & 207$\pm$36 &    0.86 &  BELLS & 1 & 0.63 &  217$\pm$38 \\
J1611+1705 &    0.477 & 1.211 & 109$\pm$23 &    0.58 &  BELLS & 1 & 1.33 &  111$\pm$23 \\
J1631+1854 &    0.408 & 1.086 & 272$\pm$14 &    1.63 &  BELLS & 1 & 2.07 &  272$\pm$14 \\
J1637+1439 &    0.391 & 0.874 & 208$\pm$30 &    0.65 &  BELLS & 1 & 0.89 &  215$\pm$31 \\
J2122+0409 &    0.626 & 1.452 & 324$\pm$56 &    1.58 &  BELLS & 1 & 1.76 &  326$\pm$56 \\
J2125+0411 &    0.363 & 0.978 & 247$\pm$17 &    1.2 &   BELLS & 1 & 1.47 &  250$\pm$17 \\
J2303+0037 &    0.458 & 0.936 & 274$\pm$31 &    1.02 &  BELLS & 1 & 1.35 &  278$\pm$31 \\
J0008-0004 &    0.44 &  1.192 & 193$\pm$36 &    1.16 &  SLACS & 1.5 &   1.71 &  197$\pm$37 \\
J0029-0055 &    0.227 & 0.931 & 229$\pm$18 &    0.96 &  SLACS & 1.5 &   2.16 &  232$\pm$18 \\
J0037-0942 &    0.196 & 0.632 & 279$\pm$10 &    1.53 &  SLACS & 1.5 &   2.19 &  283$\pm$10 \\
J0044+0113 &    0.12 &  0.196 & 266$\pm$13 &    0.79 &  SLACS & 1.5 &   2.61 &  267$\pm$13 \\
J0109+1500 &    0.294 & 0.525 & 251$\pm$19 &    0.69 &  SLACS & 1.5 &   1.38 &  259$\pm$20 \\
J0157-0056 &    0.513 & 0.924 & 295$\pm$47 &    0.79 &  SLACS & 1.5 &   1.06 &  308$\pm$49 \\
J0216-0813 &    0.332 & 0.524 & 333$\pm$23 &    1.16 &  SLACS & 1.5 &   2.67 &  335$\pm$23 \\
J0252+0039 &    0.28 &  0.982 & 164$\pm$12 &    1.04 &  SLACS & 1.5 &   1.39 &  169$\pm$12 \\
J0330-0020 &    0.351 & 1.071 & 212$\pm$21 &    1.1 &   SLACS & 1.5 &   1.2 &   220$\pm$22 \\
J0405-0455 &    0.075 & 0.81 &  160$\pm$8  &    0.8 &   SLACS & 1.5 &   1.36 &  165$\pm$8 \\
J0728+3835 &    0.206 & 0.688 & 214$\pm$11 &    1.25 &  SLACS & 1.5 &   1.78 &  219$\pm$11 \\
J0737+3216 &    0.322 & 0.581 & 338$\pm$17 &    1 & SLACS & 1.5 &   2.82 &  339$\pm$17 \\
J0808+4706 &    0.219 & 1.025 & 236$\pm$11 &    1.23 &  SLACS & 1.5 &   2.42 &  238$\pm$11 \\
J0822+2652 &    0.241 & 0.594 & 259$\pm$15 &    1.17 &  SLACS & 1.5 &   1.82 &  264$\pm$15 \\
J0841+3824 &    0.116 & 0.657 & 225$\pm$11 &    1.41 &  SLACS & 1.5 &   4.21 &  222$\pm$11 \\
J0903+4116 &    0.43 &  1.065 & 223$\pm$27 &    1.29 &  SLACS & 1.5 &   1.78 &  228$\pm$28 \\
J0912+0029 &    0.164 & 0.324 & 326$\pm$12 &    1.63 &  SLACS & 1.5 &   3.87 &  323$\pm$12 \\
J0935-0003 &    0.348 & 0.467 & 396$\pm$35 &    0.87 &  SLACS & 1.5 &   4.24 &  391$\pm$35 \\
J0936+0913 &    0.19 &  0.588 & 243$\pm$12 &    1.09 &  SLACS & 1.5 &   2.11 &  246$\pm$12 \\
J0946+1006 &    0.222 & 0.608 & 263$\pm$21 &    1.38 &  SLACS & 1.5 &   2.35 &  266$\pm$21 \\
J0956+5100 &    0.24 &  0.47 &  334$\pm$17 &    1.33 &  SLACS & 1.5 &   2.19 &  338$\pm$17 \\
J0959+0410 &    0.126 & 0.535 & 197$\pm$13 &    0.99 &  SLACS & 1.5 &   1.39 &  203$\pm$13 \\
J1016+3859 &    0.168 & 0.439 & 247$\pm$13 &    1.09 &  SLACS & 1.5 &   1.46 &  254$\pm$13 \\
J1020+1122 &    0.282 & 0.553 & 282$\pm$18 &    1.2 &   SLACS & 1.5 &   1.59 &  289$\pm$18 \\
J1023+4230 &    0.191 & 0.696 & 242$\pm$15 &    1.41 &  SLACS & 1.5 &   1.77 &  247$\pm$15 \\
J1100+5329 &    0.317 & 0.858 & 187$\pm$23 &    1.52 &  SLACS & 1.5 &   2.24 &  189$\pm$23 \\
J1106+5228 &    0.096 & 0.407 & 262$\pm$13 &    1.23 &  SLACS & 1.5 &   1.68 &  268$\pm$13 \\
J1112+0826 &    0.273 & 0.63 &  320$\pm$20 &    1.49 &  SLACS & 1.5 &   1.5 &   329$\pm$21 \\
J1134+6027 &    0.153 & 0.474 & 239$\pm$12 &    1.1 &   SLACS & 1.5 &   2.02 &  243$\pm$12 \\
J1142+1001 &    0.222 & 0.504 & 221$\pm$22 &    0.98 &  SLACS & 1.5 &   1.91 &  225$\pm$22 \\
J1143-0144 &    0.106 & 0.402 & 269$\pm$13 &    1.68 &  SLACS & 1.5 &   4.8 &   264$\pm$13 \\
J1153+4612 &    0.18 &  0.875 & 226$\pm$15 &    1.05 &  SLACS & 1.5 &   1.16 &  235$\pm$16 \\
J1204+0358 &    0.164 & 0.631 & 267$\pm$17 &    1.31 &  SLACS & 1.5 &   1.47 &  275$\pm$17 \\
J1205+4910 &    0.215 & 0.481 & 281$\pm$14 &    1.22 &  SLACS & 1.5 &   2.59 &  283$\pm$14 \\
J1213+6708 &    0.123 & 0.64 &  292$\pm$15 &    1.42 &  SLACS & 1.5 &   3.23 &  291$\pm$15 \\
J1218+0830 &    0.135 & 0.717 & 219$\pm$11 &    1.45 &  SLACS & 1.5 &   3.18 &  218$\pm$11 \\
J1250+0523 &    0.232 & 0.795 & 252$\pm$14 &    1.13 &  SLACS & 1.5 &   1.81 &  257$\pm$14 \\
J1251-0208 &    0.224 & 0.784 & 233$\pm$23 &    0.84 &  SLACS & 1.5 &   2.61 &  234$\pm$23 \\
J1330-0148 &    0.081 & 0.712 & 185$\pm$9 & 0.87 &  SLACS & 1.5 &   0.89 &  194$\pm$9 \\
J1402+6321 &    0.205 & 0.481 & 267$\pm$17 &    1.35 &  SLACS & 1.5 &   2.7 &   268$\pm$17 \\
J1403+0006 &    0.189 & 0.473 & 213$\pm$17 &    0.83 &  SLACS & 1.5 &   1.46 &  219$\pm$17 \\
J1416+5136 &    0.299 & 0.811 & 240$\pm$25 &    1.37 &  SLACS & 1.5 &   1.43 &  247$\pm$26 \\
J1430+4105 &    0.285 & 0.575 & 322$\pm$32 &    1.52 &  SLACS & 1.5 &   2.55 &  324$\pm$32 \\
J1436-0000 &    0.285 & 0.805 & 224$\pm$17 &    1.12 &  SLACS & 1.5 &   2.24 &  227$\pm$17 \\
J1451-0239 &    0.125 & 0.52 &  223$\pm$14 &    1.04 &  SLACS & 1.5 &   2.48 &  225$\pm$14 \\
J1525+3327 &    0.358 & 0.717 & 264$\pm$26 &    1.31 &  SLACS & 1.5 &   2.9 &   264$\pm$26 \\
J1531-0105 &    0.16 &  0.744 & 279$\pm$14 &    1.71 &  SLACS & 1.5 &   2.5 &   281$\pm$14 \\
J1538+5817 &    0.143 & 0.531 & 189$\pm$12 &    1  &    SLACS & 1.5 &   1.58 &  194$\pm$12 \\
J1621+3931 &    0.245 & 0.602 & 236$\pm$20 &    1.29 &  SLACS & 1.5 &   2.14 &  239$\pm$20 \\
J1627-0053 &    0.208 & 0.524 & 290$\pm$14 &    1.23 &  SLACS & 1.5 &   1.98 &  295$\pm$14 \\
J1630+4520 &    0.248 & 0.793 & 276$\pm$16 &    1.78 &  SLACS & 1.5 &   1.96 &  281$\pm$16 \\
J1636+4707 &    0.228 & 0.674 & 231$\pm$15 &    1.09 &  SLACS & 1.5 &   1.68 &  236$\pm$15 \\
J2238-0754 &    0.137 & 0.713 & 198$\pm$11 &    1.27 &  SLACS & 1.5 &   2.33 &  200$\pm$11 \\
J2300+0022 &    0.228 & 0.464 & 279$\pm$17 &    1.24 &  SLACS & 1.5 &   1.83 &  285$\pm$17 \\
J2303+1422 &    0.155 & 0.517 & 255$\pm$16 &    1.62 &  SLACS & 1.5 &   3.28 &  254$\pm$16 \\
J2321-0939 &    0.082 & 0.532 & 249$\pm$8 & 1.6 &   SLACS & 1.5 &   4.11 &  246$\pm$8 \\
J2341+0000 &    0.186 & 0.807 & 207$\pm$13 &    1.44 &  SLACS & 1.5 &   3.15 &  207$\pm$13 \\
Q0047-2808 &    0.485 & 3.595 & 229$\pm$15 &    1.34 &  LSD &   1.25 &  0.82 &  239$\pm$16 \\
CFRS03-1077 &   0.938 & 2.941 & 251$\pm$19 &    1.24 &  LSD &   1.25 &  1.6 &   256$\pm$19 \\
HST14176 &  0.81 &  3.399 & 224$\pm$15 &    1.41 &  LSD &   1.25 &  1.06 &  232$\pm$16 \\
HST15433 &  0.497 & 2.092 & 116$\pm$10 &    0.36 &  LSD &   1.25 &  0.41 &  125$\pm$11 \\
MG2016 &    1.004 & 3.263 & 328$\pm$32 &    1.56 &  LSD &   0.65 &  0.31 &  347$\pm$34 \\
J0212-0555 &    0.75 &  2.74 &  273$\pm$22 &    1.27 &  SL2S &  0.9 &  1.22 &  277$\pm$22 \\
J0213-0743 &    0.717 & 3.48 &  293$\pm$34 &    2.39 &  SL2S &  1 &  1.97 &  293$\pm$34 \\
J0214-0405 &    0.609 & 1.88 &  287$\pm$47 &    1.41 &  SL2S &  1 &  1.21 &  293$\pm$48 \\
J0217-0513 &    0.646 & 1.847 & 239$\pm$27 &    1.27 &  SL2S &  1.5 & 0.73 &  253$\pm$29 \\
J0219-0829 &    0.389 & 2.15 &  289$\pm$23 &    1.3 &   SL2S &  1 &  0.95 &  298$\pm$24 \\
J0223-0534 &    0.499 & 1.44 &  288$\pm$28 &    1.22 &  SL2S &  1 &  1.31 &  293$\pm$28 \\
J0225-0454 &    0.238 & 1.199 & 234$\pm$21 &    1.76 &  SL2S &  1 &  2.12 &  233$\pm$21 \\
J0226-0420 &    0.494 & 1.232 & 263$\pm$24 &    1.19 &  SL2S &  1 &  0.84 &  272$\pm$25 \\
J0232-0408 &    0.352 & 2.34 &  281$\pm$26 &    1.04 &  SL2S &  1 &  1.14 &  287$\pm$27 \\
J0848-0351 &    0.682 & 1.55 &  197$\pm$21 &    0.85 &  SL2S &  0.9 &  0.45 &  208$\pm$22 \\
J0849-0412 &    0.722 & 1.54 &  320$\pm$24 &    1.1 &   SL2S &  0.9 &  0.46 &  338$\pm$25 \\
J0849-0251 &    0.274 & 2.09 &  276$\pm$35 &    1.16 &  SL2S &  0.9 &  1.34 &  279$\pm$35 \\
J0850-0347 &    0.337 & 3.25 &  290$\pm$24 &    0.93 &  SL2S &  0.7 &   0.28 &  309$\pm$26 \\
J0855-0147 &    0.365 & 3.39 &  222$\pm$25 &    1.03 &  SL2S &  0.7 &   0.69 &  228$\pm$26 \\
J0855-0409 &    0.419 & 2.95 &  281$\pm$22 &    1.36 &  SL2S &  0.7 &   1.13 &  283$\pm$22 \\
J0904-0059 &    0.611 & 2.36 &  183$\pm$21 &    1.4 &   SL2S &  0.9 &  2 & 182$\pm$21 \\
J0959+0206 &    0.552 & 3.35 &  188$\pm$22 &    0.74 &  SL2S &  0.9 &  0.46 &  199$\pm$23 \\
J1359+5535 &    0.783 & 2.77 &  228$\pm$29 &    1.14 &  SL2S &  1 &  1.13 &  233$\pm$30 \\
J1404+5200 &    0.456 & 1.59 &  342$\pm$20 &    2.55 &  SL2S &  1 &  2.03 &  342$\pm$20 \\
J1405+5243 &    0.526 & 3.01 &  284$\pm$21 &    1.51 &  SL2S &  1 &  0.83 &  294$\pm$22 \\
J1406+5226 &    0.716 & 1.47 &  253$\pm$19 &    0.94 &  SL2S &  1 &  0.8 &   262$\pm$20  \\
J1411+5651 &    0.322 & 1.42 &  214$\pm$23 &    0.93 &  SL2S &  1 &  0.85 &  221$\pm$24 \\
J1420+5258 &    0.38 &  0.99 &  246$\pm$23 &    0.96 &  SL2S &  1 &  1.11 &  252$\pm$24 \\
J1420+5630 &    0.483 & 3.12 &  228$\pm$19 &    1.4 &   SL2S &  1 &  1.62 &  230$\pm$19 \\
J2203+0205 &    0.4 &   2.15 &  213$\pm$21 &    1.95 &  SL2S &  1 &  0.99 &  219$\pm$22 \\
J2205+0147 &    0.476 & 2.53 &  317$\pm$30 &    1.66 &  SL2S &  0.9 &  0.66 &  330$\pm$31 \\
J2213-0009 &    0.338 & 3.45 &  165$\pm$20 &    1.07 &  SL2S &  1 &  0.27 &  179$\pm$22 \\
J2219-0017 &    0.289 & 1.02 &  189$\pm$20 &    0.52 &  SL2S &  0.7 &  1.01 &  191$\pm$20 \\
J2220+0106 &    0.232 & 1.07 &  127$\pm$15 &    2.16 &  SL2S &  1 &  0.8 &   132$\pm$16 \\
J2221+0115 &    0.325 & 2.35 &  222$\pm$23 &    1.4 &   SL2S &  1 &  1.12 &  227$\pm$24 \\
J2222+0012 &    0.436 & 1.36 &  221$\pm$22 &    1.44 &  SL2S &  1 &  1.56 &  223$\pm$22 \\

\enddata

\end{deluxetable}

\begin{deluxetable}{lllll} \label{tab:result}
\tablewidth{0pt} \tablecaption{Dark energy ($XCDM$ model and CPL parametrization) constraints  obtained on
the full 118 strong lensing (SL) sample.}
\tablehead{ \colhead{Cosmology (Sample)} & \colhead{$w_0$} & \colhead{$w_1$} & \colhead{$\gamma_0$ }  & \colhead{$\gamma_1$}     }

\startdata
XCDM1 (SL; $\sigma_{ap}$) & $w_0=-1.45^{+0.54}_{-0.95}$    &$w_1=0$   & $\gamma_0=2.03\pm 0.06$ & $\gamma_1=0$   \\
XCDM1 (SL; $\sigma_{0}$)  & $w_0=-1.15^{+0.56}_{-1.20}$    &$w_1=0$  & $\gamma_0=2.07\pm 0.07$ & $\gamma_1=0$   \\
\hline

XCDM2 (SL; $\sigma_{ap}$) & $w_0=-1.48^{+0.54}_{-0.94}$    &$w_1=0$  & $\gamma_0=2.06\pm 0.09$ & $\gamma_1=-0.09\pm0.16$   \\
XCDM2 (SL; $\sigma_{0}$)  & $w_0=-1.35^{+0.67}_{-1.50}$    &$w_1=0$  & $\gamma_0=2.13^{+0.07}_{-0.12}$ & $\gamma_1=-0.09\pm0.17$   \\
\hline

CPL1 (SL; $\sigma_{ap}$) & $w_0=-0.15^{+1.27}_{-1.60}$    & $w_1=-6.95^{+7.25}_{-3.05}$  & $\gamma_0=2.08\pm 0.09$ & $\gamma_1=-0.09\pm 0.17$   \\
CPL1 (SL; $\sigma_{0}$)  & $w_0=-1.00^{+1.54}_{-1.95}$    & $w_1=-1.85^{+4.85}_{-6.75}$   & $\gamma_0=2.14^{+0.07}_{-0.10}$  & $\gamma_1=-0.10\pm 0.18$   \\
\hline

CPL2 (SL; $\sigma_{ap}$) & $w_0=-0.16^{+1.21}_{-1.48}$    & $w_1=-6.25^{+6.25}_{-3.75}$  & $\gamma_0=2.08$ & $\gamma_1=-0.09$   \\
CPL2 (SL; $\sigma_{0}$)  & $w_0=-1.05^{+1.43}_{-1.77}$    & $w_1=-1.65^{+4.25}_{-6.35}$   & $\gamma_0=2.14$  & $\gamma_1=-0.10$   \\
CPL2 (SN)                & $w_0=-1.00\pm0.40$             & $w_1=-0.12^{+1.58}_{-2.78}$  & $\square$      & $\square$   \\
\hline 
\enddata
\tablenotetext{a}{ In our fits we separately considered observed
velocity dispersions $\sigma_{ap}$ and corrected velocity
dispersions $\sigma_0$, $XCDM1$ corresponds to assumption of
non-evolving power-law index $\gamma$, while $XCDM2$ assumes its
evolution $\gamma(z) = \gamma_0 + \gamma_1\; z_l$. Fixed prior of
$\Omega_m=0.315$ was assumed according to the Planck data. While
fitting CPL parameters we assumed evolving lens mass density with
$\gamma_0$ and $\gamma_1$ as free parameters (CPL1) and then fixed
them at best-fit values (CPL2). For comparison fits of CPL
parameters using Union2.1 supernovae data (SN) is shown.}
\end{deluxetable}


\begin{thebibliography}{}



\bibitem[Ade et~al.(2014)]{Ade14} Ade, P.A.R., et al. [Planck Collaboration] 2014, accepted by A\&A
\bibitem[Auger et~al.(2009)]{Auger09} Auger, M. W., et al. 2009, ApJ, 105, 1099
\bibitem[Auri\`ere(1982)]{aur82} Auri\`ere, M. 1982, A\&A, 109, 301
\bibitem[Bar-Kana(1996)]{Bar-Kana96} Bar-Kana, R. 1996, ApJ, 468, 17
\bibitem[Barnab{\'e}, Spiniello \& Koopmans (2014)]{Barnabe14} Barnab{\'e}, M., Spiniello, C., \& Koopmans, L.V.E., 2014 ``Dissecting the 3D structure of elliptical galaxies with gravitational lensing and stellar kinematics'', in Eds. B.L. Ziegler, F. Combes, H. Dannerbauer, M. Verdugo, ``Galaxies in 3D across the Universe'', Proceedings of the IAU Symposium No. 309 [arXiv:1409.4197]
\bibitem[Bassett \& Kunz(2004)]{Bassett04} Bassett, B. A., \& Kunz, M. 2004, PRD, 69, 101305
\bibitem[Biesiada(2006)]{Biesiada06} Biesiada, M. 2006, PRD, 73, 023006
\bibitem[Biesiada, Pi\'{o}rkowska, \& Malec(2010)]{Biesiada10} Biesiada, M., Pi\'{o}rkowska, A., \& Malec, B. 2010, MNRAS, 406, 1055
\bibitem[Biesiada, Malec \& Pi\'{o}rkowska(2011)]{Biesiada11} Biesiada, M., Malec, B., \& Pi\'{o}rkowska, A. 2011, RAA, 11, 641
\bibitem[Bolton et~al.(2008a)]{Bolton08a} Bolton, A. S., et al. 2008a, ApJ, 682, 964
\bibitem[Brownstein et~al.(2012)]{Brownstein12} Brownstein, et al. 2012, ApJ, 744, 41
\bibitem[Canizares et al.(1978)]{can78} Canizares, C. R., Grindlay, J. E., Hiltner, W. A., Liller, W., \& McClintock, J. E. 1978, ApJ, 224, 39
\bibitem[Cao \& Zhu(2011a)]{Cao11a} Cao, S., \& Liang, N. 2011a, RAA, 11, 1199
\bibitem[Cao \& Zhu(2011b)]{Cao11b} Cao, S., \& Zhu, Z.-H. 2011b, China Series G, 54, 2260 [arXiv:1102.2750]
\bibitem[Cao et al.(2012)]{Cao12} Cao, S., et al. 2012, JCAP, 3, 16
\bibitem[Cao \& Zhu(2014)]{Cao14} Cao, S., \& Zhu, Z.-H. 2014, PRD, 90, 083006
\bibitem[Chae et al.(2002)]{Chae02} Chae, K.-H., et al. 2002, PRL, 89, 151301
\bibitem[Chae(2003)]{Chae03} Chae, K.-H. 2003, MNRAS, 346, 746
\bibitem[Chevalier \& Polarski(2001)]{Chevalier01} Chevalier, M., \& Polarski, D. 2001, IJMPD, 10, 213
\bibitem[Collett et al. (2014)]{panglos} Collett T.E., et al. 2013, MNRAS, 432, 679-692  [arXiv:1303.6564]
\bibitem[Djorgovski \& King(1984)]{djo84} Djorgovski, S., \& King, I. R. 1984, ApJL, 277, L49
\bibitem[Eisenstein et al.(2011)]{Eisenstein11} Eisenstein, D. J., et al. 2011, AJ, 142, 72 [arXiv:1101.1529]
\bibitem[Falco, Gorenstein \& Shapiro (1985)]{Falco85} Falco, E. E., Gorenstein, M. V., Shapiro, I. I. 1985, ApJ, 289, L1
\bibitem[Futamase \& Yoshida(2001)]{Futamase01} Futamase, T., \& Yoshida, S. 2001, Prog. Theor. Phys., 105, 887
\bibitem[Gavazzi et al. (2014)]{RingFinder} Gavazzi, R., Marshall, P.J., Treu, T. \& Sonnenfeld, A., ``RingFinder: automated detection of galaxy-scale gravitational lenses in ground-based multi-filter imaging data'', (2014) ApJ accepted [arXiv:1403:1041]
\bibitem[Gilmore \& Natarayan(2009)]{Gilmore09} Gilmore, J., \& Natarayan, P. 2009, MNRAS, 396, 354
\bibitem[Grillo et~al.(2008)]{Grillo08} Grillo, C., Lombardi, M., \& Bertin, G. 2008, A\&A, 477, 397
\bibitem[Holanda, Lima \& Ribeiro(2010)]{HLB10} Holanda, R. F. L., Lima, J. A. S., \& Ribeiro, M. B., 2010, ApJ, 722, L233
\bibitem[Jaroszy\'{n}ski \& Kostrzewa-Rutkowska(2012)]{Jaroszyki12} Jaroszy\'{n}ski, M., \& Kostrzewa-Rutkowska, Z. 2012, MNRAS, 424, 325
\bibitem[J{\o}rgensen et~al.(1995a)]{Jorgensen95a} J{\o}rgensen, I., Franx, M. \& Kj{\ae}rgard, P. 1995a, MNRAS, 273, 1097
\bibitem[J{\o}rgensen et~al.(1995b)]{Jorgensen95b} J{\o}rgensen, I., Franx, M. \& Kj{\ae}rgard, P. 1995b, MNRAS, 276, 1341
\bibitem[Jullo et~al.(2010)]{Jullo10} Jullo, E., et al. 2010, Science, 329, 924
\bibitem[Keeton, Kochanek \& Seljak(1997)]{Keeton97} Keeton, C. R., Kochanek, C. S., \& Seljak, U. 1997, ApJ, 487, 42
\bibitem[Kochanek \& Apostolakis(1988)]{Kochanek88} Kochanek, C. S., \& Apostolakis, J. 1988, MNRAS, 235, 1073
\bibitem[Kochanek et al.(2000)]{Kochanek00} Kochanek C., et al. 2000, ApJ, 543, 131
\bibitem[Koopmans \& Treu(2002)]{Koopmans02} Koopmans, L.V.E, \& Treu, T. 2002, ApJ, 583, 606
\bibitem[Koopmans et~al.(2005)]{Koopmans05} Koopmans L.V.E. 2005, Proceedings of XXIst IAP Colloquium, ``Mass Profiles \& Shapes of Cosmological Structures'' (Paris, 4-9 July 2005), eds G. A. Mamon, F. Combes, C. Deffayet, B. Fort (Paris: EDP Sciences)  [astro-ph/0511121]
\bibitem[Koopmans et~al.(2006)]{Koopmans06} Koopmans, L.V.E., et al. 2006, ApJ, 649, 599
\bibitem[Koopmans et~al.(2009)]{Koopmans09} Koopmans, L.V.E., et al. 2009, ApJ, 703, L51
\bibitem[Lewis \& Bridle(2002)]{Lewis02} Lewis, A., \& Bridle, S. 2002, PRD, 66, 103
\bibitem[Linder(2003)]{Linder03} Linder, E. V. 2003, PRD, 68, 083503
\bibitem[Linder(2004)]{Linder04} Linder, E. V. 2004, PRD, 70, 043534
\bibitem[Linder(2011)]{Linder11} Linder, E. V. 2011, PRD, 84, 123529
\bibitem[Meneghetti et~al.(2005)]{Meneghetti05} Meneghetti, M., et al. 2005, A\&A, 442, 413
\bibitem[Oguri et al.(2012)]{Oguri12} Oguri, M., et al. 2012, AJ, 143, 120
\bibitem[Paczy{\'n}ski \& G{\'o}rski(1981)]{Paczynski81} Paczy{\'n}ski, B., \& G{\'o}rski, K. 1981, ApJ, 248, L101
\bibitem[Perlmutter et~al.(1999)]{Perlmutter99} Perlmutter, S., et~al. 1999, ApJ, 517, 565
\bibitem[Piorkowska et al.(2013)]{Piorkowska13} Pi\'orkowska A., et al. 2013, Acta Phys. Polon. B, 44, 2397
\bibitem[Ratnatunga et~al.(1999)]{Ratnatunga99} Ratnatunga, K. U., Griffiths, R. E., \& Ostrander, E. J. 1999, AJ, 117, 2010
\bibitem[Ratra \& Peebles(1988)]{Ratra88} Ratra, B., Peebles, P. E. J., 1988, PRD, 37, 3406
\bibitem[Refsdal(1964)]{Refsdal64} Refsdal, S. 1964, MNRAS, 128, 295
\bibitem[Riess et~al.(1998)]{Riess98} Riess, A.~G., et~al. 1998, AJ, 116, 1009
\bibitem[Ruff et~al.(2011)]{Ruff11} Ruff, A., et al. 2011, ApJ, 727, 96
\bibitem[Rusin, Kochanek \& Keeton(2003)]{Rusin03} Rusin, D., Kochanek, C. S., \& Keeton, C. R. 2003, ApJ, 595, 29
\bibitem[Schneider et~al.(1992)]{Schneider92} Schneider, P., Ehlers, J., \& Falco, E.~E. 1992, Gravitational Lenses
\bibitem[Schneider \& Sluse (2013)]{SchneiderSluse} Schneider, P. \& Sluse, D. 2013, A\&A, 559, A37
\bibitem[Sereno(2002)]{Sereno02} Sereno, M. 2002, A\&A, 393, 757
\bibitem[Shu et al. (2014)]{Shu14} Shu, Y. et al., ``The Sloan Lens ACS Survey, XII. Extending Strong Lensing to Lower Masses'', (2014) ApJ submitted [arXiV: 1407.2240]
\bibitem[Sonnenfeld et~al.(2013a)]{Sonnenfeld13a} Sonnenfeld, A., Gavazzi, R., Suyu, S.H., Treu, T., Marshall, P.J. 2013a, ApJ, 777, 97 [arXiV:1307.4764]
\bibitem[Sonnenfeld et~al.(2013b)]{Sonnenfeld13b} Sonnenfeld, A., Treu, T., Gavazzi, R., Suyu, S.H., Marshall, P.J., Auger, M.W., Nipoti, C., 2013b, ApJ, 777, 98 [arXiv:1307.4759v1]
\bibitem[Suyu et al.(2010)]{Suyu10} Suyu, S. H, et al. 2010, ApJ, 711, 201
\bibitem[Suzuki et al.(2012)]{Suzuki12} Suzuki, N., et al. 2012, ApJ, 746, 85
\bibitem[Treu \& Koopmans(2002)]{Treu02} Treu, T., \& Koopmans, L.V.E. 2002, ApJ, 575, 87
\bibitem[Treu \& Koopmans(2004)]{Treu04} Treu, T.,\& Koopmans, L.V.E. 2004, ApJ, 611, 739
\bibitem[Treu et~al.(2006)]{Treu06} Treu, T., et al. 2006, ApJ, 650, 1219
\bibitem[Uzan et al.(2004)]{Uzan04} Uzan, J. P., Aghanim, N., \& Mellier, Y. 2004, PRD, 70, 083533
\bibitem[Wambsganss, Bode \& Ostriker(2005)]{Wambsganss05} Wambsganss, J., Bode, P., \& Ostriker, J. P., 2005, ApJ, 635, L1

\end{thebibliography}
\end{document}